\documentclass[tightenlines,aps,floatfix,preprint,
nofootinbib]{revtex4}
\usepackage{float,axodraw,epsfig}

\newcommand{\ba}{\begin{eqnarray}}
\newcommand{\ea}{\end{eqnarray}}

\newcommand{\ep}{\epsilon}

\newcommand{\be}{\begin{equation}}
\newcommand{\ee}{\end{equation}}

\def\hname{{\sf ~FEHiP}}
\begin{document}

\title{
\hfill{\normalsize\vbox{%
\hbox{\rm UH-511-1066-05}
}}
\vspace*{0.5cm}

Fully differential Higgs boson production and the di-photon signal through next-to-next-to-leading order
}

\author{
Charalampos Anastasiou\thanks{e-mail:babis@phys.ethz.ch}} 
\affiliation{
          Institute for Theoretical Physics,\\ 
          ETH, 8093 Zurich, Switzerland}
\author{Kirill Melnikov
        \thanks{e-mail: kirill@phys.hawaii.edu}}
\affiliation{Department of Physics and Astronomy,
          University of Hawaii,\\ 2505 Correa Rd., Honolulu, Hawaii 96822}  
\author{Frank Petriello\thanks{frankp@pha.jhu.edu}}
\affiliation{
Department of Physics, Johns Hopkins University, \\
3400 North Charles St., Baltimore, MD 21218
}

\begin{abstract}
We describe a calculation of the 
fully differential cross section for Higgs boson production 
in the gluon fusion channel through next-to-next-to-leading order (NNLO)
in perturbative QCD. The decay of the Higgs 
boson into two photons is included.  Technical aspects of 
the computation are discussed in detail. 
The implementation of the calculation 
into a numerical code, called \hname, is described. The NNLO $K$-factors 
for completely realistic photon acceptances and isolation cuts, including those employed by the 
ATLAS and CMS collaborations, are computed. 
We study various distributions of the photons from Higgs decay through NNLO.

\end{abstract}

\maketitle

\section{Introduction}

Perturbative calculations in quantum field theory have been 
performed since the birth of QED in the 1930s.  Enormous progress 
has been made since this early era, and the tools used to obtain 
these results have improved dramatically.  Much of this progress 
has been spurred by the constantly increasing demands of 
the high energy experimental program.  Future experiments 
such as the LHC and the Next Linear Collider are now demanding 
perturbative calculations to next-to-next-to-leading order (NNLO) 
in the relevant coupling constants.  These results must also be 
directly applicable to experimental measurements, which requires 
calculational algorithms flexible enough to allow arbitrary 
constraints on the final state of reactions.  

For a long time, virtual corrections hindered the progress of 
perturbative calculations.  With the advent of computer algebra and the realization that 
loop integrals satisfy simple identities that follow from 
their generalized hypergeometric nature \cite{tkachov}, virtual corrections for 
processes with a small number of kinematic invariants became relatively 
straightforward \cite{loop1,loop2,loop3,laporta}.  However, physical 
results require the inclusion of real emission corrections, in which 
additional massless partons are emitted into the final state.  It was 
recognized early in the history of perturbative calculations, 
first in QED and then in QCD, that such emissions lead to infrared and collinear singularities 
when the initial or final state configurations become degenerate.  The 
Bloch-Nordsieck and Kinoshita-Lee-Nauenberg theorems \cite{Bloch:1937pw,Kinoshita:1962ur,Lee:1964is} 
show that both virtual corrections and real emissions must be included 
to obtain a physically meaningful result, because divergences only 
cancel when these components are combined.  The methods cited above  
rely upon special features of loop integrals.  They can  be 
used to obtain virtual corrections, or inclusive cross sections related to 
virtual corrections through the optical theorem, such as 
the total hadroproduction cross section in $e^+e^-$ collisions.  
Recently, these methods have also been extended to phase-space integrals for total cross sections and 
simple kinematic distributions~\cite{Anastasiou:2002yz,Anastasiou:2002wq,Anastasiou:2002qz,Anastasiou:2003yy,
Anastasiou:2003ds,Melnikov:2004bm,Mitov:2004du}. However, observables where the phase space 
is non-trivially constrained cannot be obtained using these techniques.  
Unfortunately, these are exactly the quantities required by experiment; 
because of final-state cuts, inclusive results are of limited use.

Virtual corrections possess a simple mathematical structure, and 
the extraction of their singularities proceeds in an 
observable-independent fashion.  With real corrections, 
different kinematic cuts are imposed on the final state for different 
observables.  Since we would like to perform calculations valid 
for arbitrary cuts, the extraction of singularities from real emission 
corrections should be performed in the presence of 
an unspecified ``measurement function''.  
The factorization of soft and collinear emissions renders such an 
extraction possible.  The resulting cancellation of singularities requires 
that the measurement function allows only ``infrared safe'' observables that 
can be computed in perturbation theory~\cite{sterman}.  

An efficient algorithm for extracting  singularities at NLO in 
perturbative QCD was constructed in~\cite{Catani:1996vz,Catani:2002hc}, 
advancing
earlier work on the 
subject~\cite{nloreal}. 
This ``dipole-subtraction'' algorithm identifies universal soft and 
collinear counterterms, called dipoles, that can be subtracted from 
the real emission contribution to an arbitrary process
to make it finite.  The dipoles can then be 
analytically integrated over the restricted phase space, since the measurement 
function for infrared safe observables collapses to a measurement 
function of lower multiplicity in the soft and collinear limits.  After 
integration, the dipoles cancel the infrared divergences in the virtual 
corrections, leading to a finite result.

For processes where the perturbative corrections are large, or 
control of the theoretical error is crucial, perturbative 
calculations must be extended to NNLO.  Either an extension of the 
dipole formalism to NNLO or the development of an alternative approach 
is needed to enable these computations.  Significant effort has been 
devoted to a generalization of the dipole subtraction method  
\cite{Kosower:2002su,Kosower:2003cz,Weinzierl:2003fx,Weinzierl:2003ra,Gehrmann-DeRidder:2004tv,Gehrmann-DeRidder:2003bm,Gehrmann-DeRidder:2004xe,Kilgore:2004ty,Frixione:2004is}.  This extension has remained elusive.  
Although the infrared behavior of 
loop amplitudes~\cite{catani,tejeda} and the infrared 
limits of real emission corrections are universal at NNLO \cite{amplfactor}, 
it becomes much more difficult to disentangle 
the singularities of two unresolved emissions, and to construct 
process-independent counterterms.

We have recently developed an alternative approach to the 
problem of real radiation at NNLO~\cite{sector}.  Our method differs 
conceptually from the dipole subtraction approach in two important ways: 
the finding of singular phase-space regions is completely automated; the 
cancellation of the $1/\epsilon$ poles which describe divergences in 
dimensional regularization is performed numerically, and 
no analytic integrations are required.  These features guarantee that it can 
be used to extract and cancel singularities at {\it any} order in 
perturbation theory, at least in principle.  The primary complication that 
occurs at higher orders is the presence of overlapping singularities.  
These can be disentangled using sector decomposition~\cite{Binoth:2000ps,hepp,Roth:1996pd}.  
Existing symbolic manipulation programs, such as MAPLE or MATHEMATICA, 
provide a suitable framework in which to program the  algorithm.  The 
extraction and cancellation of singularities is thus achieved in a 
completely automated fashion, with little human intervention.

We have used this approach to perform two fully differential calculations 
through NNLO in QCD: the ${\cal O}(\alpha_s^2)$ correction to $e^+e^- \to 2~{\rm jets}$ 
\cite{sector,Anastasiou:2004qd}, and the Higgs boson hadroproduction cross section 
through gluon fusion~\cite{higgsdiff}. 
Both calculations permit arbitrary cuts on the 
final states and can easily be extended to include decays of the final state particles.  
These successful applications demonstrate the vitality of our approach and its potential 
relevance for other problems of phenomenological importance; we therefore believe 
it is important to describe it in a simple, pedagogical fashion.  We attempt to do so
in this manuscript.  Our discussion will be centered around  
the calculation of the Higgs hadroproduction cross section completed recently~\cite{higgsdiff}.  We will 
extend this result to include the decay of the Higgs through the channel 
$H \rightarrow \gamma\gamma$.  We therefore pursue two goals in this 
paper: a thorough discussion of the analytic aspects of our calculation and its 
implementation into a numerical code, and a presentation of phenomenological results 
for $pp \rightarrow H+X \rightarrow \gamma\gamma +X$ through 
NNLO in QCD.

This paper is organized as follows.   In the next Section we 
describe phenomenological issues relevant for Higgs boson hadroproduction.  
In Section \ref{sec:notation} we introduce our notation, and describe 
the basic setup of our calculation.  In Section \ref{sec:soloHiggs} we discuss Higgs production in association 
with up to one parton.  Section \ref{sect:collinear} is devoted 
to the calculation of the collinear counterterms.
The phase-space parameterization is a crucial element of our approach; we 
describe details regarding the choice of parameterizations for 
Higgs hadroproduction in Section \ref{sect:phspacepar}.
We discuss how to handle the various 
forms of singularities which appear in NNLO computations 
in the following Section.  In Section \ref{sect:top}
we discuss how 
the matrix elements that appear in the 
double real emission contribution for Higgs hadroproduction 
are treated in our calculation.
  We then 
present phenomenological results for $pp \rightarrow H+X$ and 
$pp \rightarrow H+X \rightarrow \gamma\gamma +X$ through NNLO in 
perturbative QCD.  We next describe the implementation of our results into the numerical code \hname.
We conclude with a discussion of the advantages and disadvantages 
of our approach, and identify directions for future work.

\section{Higgs boson production at hadron colliders}

We describe here the aspects of Higgs phenomenology at the LHC 
needed for our calculation.  There are several mechanisms for 
Higgs production at hadron colliders (see \cite{Carena:2002es} for a review).  
For most Higgs masses, gluon fusion through a top quark loop, 
$gg \to H$, is the dominant production mechanism.  
For Higgs masses in the range preferred 
by the global fit to the precision electroweak data~\cite{ewkprecision}, 
$m_h \approx 110-130$ GeV, the gluon fusion cross section is 
approximately $60~{\rm pb}$.  The 
Higgs branching fraction into two photons is larger than 
$10^{-3}$ in this mass range \cite{prod,dec}.  The search strategy \cite{search} 
for the Higgs signal is then to look for events with two isolated 
photons and reconstruct the mass of the Higgs 
boson by studying their invariant mass distribution.  The photons are 
required to have transverse momenta $p_\bot^{(1)} \ge 40~{\rm GeV}$ and 
$p_\bot^{(2)} \ge 25~{\rm GeV}$; they must also be produced in the central 
rapidity region $|\eta| <2.5$~\cite{search}. The major irreducible background 
to two photon events is direct (prompt) 
di-photon production in hadronic collisions \cite{diphot};  
a $1\%$ photon energy resolution is needed to distinguish the $H \to \gamma \gamma$ 
signal over the background.  Typically, an isolation cut is also imposed 
upon the photons; this suppresses photons from the decays of large 
$p_\bot$ hadrons, such as  $\pi^0 \to \gamma \gamma$, and from the poorly 
known fragmentation production of prompt photons.  Several 
possible isolation cuts have been 
proposed~\cite{lance1,smoothcone,difox,search}.  
The simplest possibility is to require that a photon candidate does not have 
additional transverse energy  $E_{\rm T,min}$ deposited within the region 
$R_{\rm is} = \sqrt{(\eta-\eta_\gamma)^2 +  (\phi-\phi_\gamma)^2}$
in the $(\eta,\phi)$ plane.  Typical values 
used in previous studies are $E_{\rm T,min} = 4-15~{\rm GeV}$ 
and $R_{\rm is} = 0.4$. 

The Higgs production cross section 
receives large perturbative corrections and depends strongly 
on the renormalization and factorization scales.
For example, for $m_h = 100-130~{\rm GeV}$, the cross section 
for $pp \to H+X$ at $\sqrt{s}=14~{\rm TeV}$ 
increases by a factor $1.5-1.7$ when the NLO QCD corrections are 
included \cite{dawson,spira}. 
The residual 
scale dependence at NLO is approximately thirty percent.
This peculiar behavior of the perturbative series 
 motivated several NNLO calculations of the inclusive Higgs production cross 
section~\cite{Harlander:2001is,Harlander:2002wh,Catani:2001ic,
Anastasiou:2002yz,Ravindran:2003um}. 
These studies found no breakdown of the perturbative expansion; while the NNLO effects are sizable, 
they are smaller than the NLO ones.  The NNLO 
cross section is also much more stable against variations of the 
renormalization and the factorization scales.  These results were 
confirmed \cite{Catani:2003zt} in the framework of threshold 
resummation, which exploits the fact that because of the large value 
of the gluon density at small Bjorken $x$, 
the Higgs production cross section
at the LHC is dominated by the partonic threshold, i.e., the $z \to 1$ 
region, where $z = m_h^2/s_{\rm part}$. The terms that 
are singular in this  limit can be systematically resummed 
to all orders in $\alpha_s$, and the results compared 
to the complete NNLO calculation.  The two approaches agree well, indicating that the 
uncalculated 
higher order effects are likely to be within the uncertainty assigned 
to the NNLO result.

Much is also known about less 
inclusive quantities for Higgs boson production. 
The NLO $p_\bot$ and rapidity 
distributions for Higgs boson production at high $p_\bot$ are computed in~\cite{Ravindran:2002dc}. 
In addition, the rapidity distribution of the 
Higgs boson has been computed through NLO \cite{Anastasiou:2002qz,lance1}.  
The $p_\bot$ distribution of the Higgs boson has been 
investigated using various resummation formalisms by different groups~\cite{group1,group2,group3,group4}.  
Monte Carlo event generators accurate through NLO for 
the ${\rm Higgs} + {\rm jet}$ process have been 
published~\cite{deFlorian:1999zd,Campbell:2000bg}. Higgs production is also included in existing shower Monte Carlo 
event generators, such as PYTHIA and HERWIG, and in the 
MC@NLO event generator that correctly combines single 
hard gluon emissions with the HERWIG parton shower \cite{mcatnlo}.  
The di-photon invariant mass 
distribution in $pp$ collisions, including both the signal 
$pp \to H +X \to \gamma \gamma +X$ and the prompt photon 
background $pp \to \gamma \gamma$, can be found in 
\cite{difox,lance1,lance2}. Ref.~\cite{difox} presents 
the partonic level event generator DIPHOX, where both direct 
and fragmentation components of the prompt photon production 
are computed through NLO in perturbative QCD. 
In \cite{lance1}, the QCD corrections to the $gg \to \gamma \gamma$ 
channel of diphoton production are computed; combined 
with DIPHOX, this presents the most up-to-date analysis
of the two photon background to Higgs production and enables
a careful analysis of the signal-to-background ratio as a function 
of the isolation cuts. In \cite{lance2}, the interference between the 
signal and background is shown to be negligible.

Clearly, a substantial amount is known about Higgs hadroproduction; 
unfortunately, only the inclusive cross section is known through NNLO.  Since the 
NNLO corrections are large, it is desirable to also know 
differential quantities to this order.  This is necessary to 
compute the $K$-factor, $K_{\rm NNLO} = \sigma_{\rm NNLO}/\sigma_{\rm LO}$, 
that corresponds to realistic experimental acceptances with cuts 
on the photons and jets in the final state.  Although the 
kinematics of the $H \rightarrow \gamma\gamma$ decay is not altered 
by higher order QCD effects\footnote{Note that this
statement is violated by the decay of the Higgs boson into 
two gluons and two photons, $H \to gg \gamma \gamma$.}, 
the QCD corrections to the production 
process change the kinematics of the produced Higgs boson 
and lead to modifications in the kinematics of the produced 
photons.  In order to compute the relevant acceptance
through NNLO, the Higgs boson production cross section must 
be known at the differential level.  We begin our description of 
this calculation in the following Section.

\section{Notation and Setup}
\label{sec:notation}

We study the production of a Higgs boson with momentum $p_h$ 
in the collision of two hadrons, $h_1$, $h_2$, carrying momenta 
$P_1, P_2$:
\begin{equation}
h_1(P_1) + h_2(P_2) \to H(p_h) + X. 
\end{equation}
Within the framework of QCD factorization, the cross section for this process 
can be written as an integral over hard scattering cross sections 
$\sigma_{ij}$ for the production of the Higgs boson from the quarks 
and gluons, multiplied by parton densities describing the distribution 
of these partons inside the colliding hadrons:
\begin{equation}
\label{eq:xsection_bare}
\sigma = \sum_{ij} 
\int_0^1 dx_1 dx_2 
f_i^{\left( h_1 \right)}(x_1)
f_j^{\left( h_2 \right)}(x_2)
\sigma_{ij \to H+X}(x_1, x_2).
\end{equation}
The sum is over the parton flavors $i,j$ in the hadrons $h_1,h_2$, 
and $f_i^{(h_1)},f_j^{(h_2)}$ are the corresponding parton densities.
The initial-state partons $i,j$ for the hard scattering partonic process 
carry momenta $p_1 = x_1 P_1$ and
$p_2 =x_2 P_2$. The partonic cross sections for the 
processes $i + j \to H +X$, can be calculated  perturbatively; here, 
we compute them through ${\cal O}(\alpha_s^4)$ in the strong coupling 
expansion:
\begin{equation}
\sigma_{ij \to H+X} = \alpha_s^2\left[\sigma_{ij}^{(0)} + 
\frac{\alpha_s}{\pi} \sigma_{ij}^{(1)} +
\left(\frac{\alpha_s}{\pi}\right)^2 \sigma_{ij}^{(2)} 
\right] + {\cal O}(\alpha_s^5).
\end{equation}

The partonic cross sections $\sigma_{ij}$ contain divergences arising 
from intial-state collinear radiation; these are removed by recasting 
the parton-densities in the $\overline{\rm MS}$-factorization scheme.
The hadronic cross section of Eq.~\ref{eq:xsection_bare} is computed in 
terms of finite parton densities $\tilde{f}$ and finite partonic cross sections 
$\hat{\sigma}$,
\begin{equation}
\label{eq:xsection}
\sigma = \sum_{ij} 
\int_0^1 dx_1 dx_2 
\tilde{f}_i^{\left( h_1 \right)}(x_1)
\tilde{f}_j^{\left( h_2 \right)}(x_2)
{\hat{\sigma}}_{ij \to H+X}(x_1, x_2).
\end{equation}
The finite and ``bare'' parton densities are related via
\begin{equation}
\tilde{f}_i^{(h)} = \sum_j f_j^{(h)}  \otimes \Gamma_{ij},
\end{equation}  
where we have introduced the convolution integral 
\begin{equation}
\label{eq:dressbare}
(f \otimes g) (x) = \int_0^1 dydz f(y)g(z)\delta(x-yz).
\end{equation}
The functions $\Gamma_{ij}$ are given in the $\overline{ {\rm MS}}$ scheme 
by 
\begin{eqnarray}
\Gamma_{ij}(x)& =& \delta_{ij} \delta(1-x) - 
\frac{\alpha_s}{\pi} \frac{P_{ij}^{(0)}}{\epsilon} \nonumber \\
&& \hspace{-1.5cm}
+\left( \frac{\alpha_s}{\pi} \right)^2
\left\{ 
\frac{1}{2\epsilon^2} \left[ 
\sum_k \left(P_{ik}^{(0)} \otimes P_{kj}^{(0)}\right)(x)
+ \beta_0 P_{ij}^{(0)}
\right]
-\frac{1}{2\epsilon} P_{ij}^{(1)}(x)
\right\} 
+ {\cal O}\left( \alpha_s^3 \right)
\end{eqnarray}
where the Altarelli-Parisi kernels $P_{ij}^{(n)}$ 
can be found in~\cite{splitting1,splitting2}.  We note that the complete 
NNLO corrections to these kernels have recently been computed~\cite{splitting3}.
$\epsilon = (4-d)/2$ is 
the usual dimensional regularization parameter; all calculations in 
this paper are performed using this regularization scheme.
Substituting Eq.~(\ref{eq:dressbare}) into Eq.~(\ref{eq:xsection}) and 
comparing with Eq.~(\ref{eq:xsection_bare}) we find
\begin{equation}
\label{eq:sigma_convol}
\sigma_{ij} = \sum_{kl} 
\int_0^1 dy_1 dy_2
\Gamma_{ik}(y_1) 
\Gamma_{jl}(y_2)
\hat{\sigma}_{kl}(x_1 y_1, x_2 y_2).
\end{equation} 
We compute the finite partonic cross sections $\hat{\sigma}_{ij}$
by expanding 
\begin{equation}
\hat{\sigma}_{ij} = \alpha_s^2\left[\hat{\sigma}_{ij}^{(0)} + 
\frac{\alpha_s}{\pi} \hat{\sigma}_{ij}^{(1)} +
\left(\frac{\alpha_s}{\pi}\right)^2 \hat{\sigma}_{ij}^{(2)} 
\right] + {\cal O}(\alpha_s^5),
\end{equation}
and solving Eq.~(\ref{eq:sigma_convol}) in terms of the coefficients
$\hat{\sigma}_{ij}^{(n)}$ order-by-order in the strong coupling expansion. 
In this procedure, we need to consider the convolution integrals of the 
partonic cross sections with the Altarelli-Parisi kernels at each order in
the perturbative expansion; this will be discussed in detail in a later Section.  

We now discuss the Lagrangian which describes Higgs boson production.  As 
mentioned in the previous Section, we consider the gluon fusion mechanism for 
Higgs production.  The Higgs coupling to two gluons is induced by a top quark loop \cite{prod}; 
if there are other heavy quark doublets that acquire mass from the 
Higgs mechanism, they might also give a substantial contribution to 
the effective $Hgg$ coupling.  We focus here on a light Standard Model 
Higgs boson whose mass is smaller than 
twice the mass of the top quark: $m_h \le 2m_t\approx 350~{\rm GeV}$.  
The interaction of the Higgs boson
with two gluons can then be described 
by a point-like vertex \cite{prod}; this is formalized by introducing 
the effective Lagrangian
\begin{equation}
\label{eq:lagrangian}
{\cal L} = \frac{1}{4v} C_1 Z_1 G^a_{\mu\nu} G^{a \mu \nu} H,
\end{equation}
where  $G^a_{\mu\nu}$ is the gluon field strength tensor, 
$H$ is the Higgs field, 
and $v \simeq 246$ GeV is the Higgs boson vacuum expectation value. 
The Wilson coefficient $C_1$ and the 
renormalization factor $Z_1$, defined in the $\overline{{\rm MS}}$ 
scheme, are~\cite{koeffc1} 
\begin{eqnarray}
&& C_1 = -\frac{1}{3\pi} \left\{
1+ 
\frac{11}{4} \frac{\alpha_s}{\pi} 
+ \left( \frac{\alpha_s}{\pi}  \right)^2 
\left[
\frac{2777}{288} + \frac{19}{16} L_t +
n_f \left( 
-\frac{67}{96} + \frac{1}{3}L_t
\right)
\right] + 
{\cal O}\left( \alpha_s^3 \right)
\right\},
\nonumber \\
&& Z_1 = 1 -  \frac{\alpha_s}{\pi} \frac{\beta_0}{\epsilon}
+ \left(  \frac{\alpha_s}{\pi} \right)^2
\left[ 
\frac{\beta_0^2}{\epsilon^2} - \frac{\beta_1}{\epsilon}
\right] + 
{\cal O}\left( \alpha_s^3 \right),
\end{eqnarray}
where  $\alpha_s = \alpha_s(\mu)$ is the QCD $\overline{{\rm MS}}$
coupling constant, defined in the theory with $n_f=5$ flavors, and
\begin{equation}
L_t = \log\left(\frac{\mu^2}{m_{top}^2} \right),
\quad
\beta_0 = \frac{11}{4} - \frac{n_f}{6},
\quad
\beta_1 = \frac{51}{8} - \frac{19}{24}n_f.
\end{equation}
The Feynman rules for the $ggH$ vertex follow from the 
effective Lagrangian in Eq.(\ref{eq:lagrangian}).  
The Higgs boson couplings to light fermions are neglected. The leading 
order partonic process is then $gg \to H$; at higher orders 
in perturbation theory, other partonic processes also contribute.

At the LHC, the energy of hadronic collisions is much larger 
than the Higgs mass; it is therefore not obvious that 
the approximation of a point-like Higgs coupling to two gluons 
is sufficiently accurate.  However, Higgs bosons with masses 
$m_h \sim 100~{\rm GeV}$ are predominantly produced close 
to the threshold of the partonic collision,
with an average transverse 
momentum of tens of GeV.  The kinematic
invariants in the partonic $gg \to H+X$ process never become 
large enough 
to resolve the top quark loop, unless the large $p_\bot$ region 
is specifically probed.  Since the contribution from this region is 
negligibly small, the point-like approximation is valid.

We now discuss what partonic contributions are required to compute the 
differential cross section for Higgs hadroproduction at NNLO.
At the partonic level, the leading order process 
is $gg \to H$. At NLO, three other partonic processes appear:
$gg \to Hg$, $qg \to Hq$ and $q \bar q \to H g$. At NNLO, we must 
compute: i) two-loop virtual corrections 
to $gg \to H$; ii) one-loop virtual corrections 
to $gg \to Hg$, $qg \to Hq$ and $q\bar q \to Hg$; 
iii) inelastic processes with two partons in the final 
state: $gg \to Hgg$, $gg \to H q \bar q$, 
$qg \to Hqg$, $q \bar q \to H gg$, $q \bar q \to H q \bar q$, and $q_i q_j \to H q_i q_j$.  
Each of these contributions has the generic form
\begin{equation}
\label{eq:generic_csection}
\sigma = \int d\Pi_l 
\left| {{\cal M}_{ij\to H+l {\rm partons}}} \right|
F_J(x_1,x_2,p_1,p_2,\{q_l\}), 
\end{equation}  
where $d\Pi_l$ denotes the integration over the Higgs phase-space and 
the phase-space of $l$ additional partons in the final space, 
${\cal M}_{ij}$ includes both the matrix elements and any required 
loop integrations, and $F_J$ describes the observable under consideration.

The loop integrations are universal for all observables, and can be performed 
with well established methods.  We use integration-by-parts identities and recurrence 
relations \cite{tkachov}
to calculate the virtual corrections to both the LO and NLO 
processes. The recurrence relations are solved using the algorithm described in \cite{laporta} and implemented in \cite{air}. 
The resulting master integrals are then evaluated
directly. This is straightforward for the two-loop 
virtual corrections, since the $l=0$ phase-space 
integration just gives $\delta(m_{h}^2-(p_1+p_2)^2)$; for the virtual corrections 
to the NLO processes, the resulting master integrals have 
to be integrated over the $l=1$ particle phase space. 
This can produce additional singularities.  Care must be 
taken to assure that all singularities are extracted properly.  
However, because the two-particle phase-space 
is simple, the extraction of singularities is easy 
and proceeds along the lines discussed in \cite{Anastasiou:2003ds}.
 Hence, dealing with either two- or one-loop virtual corrections 
to Higgs hadroproduction is straightforward; we will discuss these briefly 
in the following Section. 

The situation is drastically different for the double real 
emission channels.  As discussed in the Introduction, efficient
 extraction of infrared and collinear singularities
from this component is still an open issue.  It is known how to compute analytically the phase-integrals for 
the total 
cross section~\cite{Harlander:2002wh,Anastasiou:2002yz,Ravindran:2003um}, 
where we must set $F_J=1$.  It is also possible to compute 
analytically simple kinematic distributions where the measurement function 
takes a simple form. For example, to compute the 
rapidity $Y$ distribution of the Higgs boson in the frame of the two hadrons 
we must insert 
\begin{equation}
F_J = \delta \left(
Y - \frac{1}{2} \log\left[ \frac{p_h^0 + p_h^z}{ p_h^0 - p_h^z}\right]
\right) =
\delta \left(
Y - \frac{1}{2} \log\left[ \frac{x_1 p_2\cdot p_h}{ x_2 p_1 \cdot p_h}\right]
\right),
\end{equation}
where the $z$-axis is the beam axis.  Phase-space integrations of this type, 
where the measurement function can be written as a delta-function constraining 
a covariant quantity, can be mapped to loop integrals and solved using the 
techniques discussed in the previous paragraph \cite{Anastasiou:2003yy,Anastasiou:2003ds}.
The rapidity distribution has been computed analytically 
for electroweak gauge boson production at hadron 
colliders~\cite{Anastasiou:2003yy,Anastasiou:2003ds}; these computations 
require similar phase-space integrations as for Higgs boson production. 
However, the measurement function $F_J$ can take very complicated forms, 
which are unsuitable for an analytic evaluation of the cross section, 
if additional components of the Higgs boson momentum or the final-state 
partonic momenta are probed, a jet finding algorithm is applied, or the decay 
of the Higgs boson with all relevant experimental cuts is included.

The difficulties related to the evaluation of the double real emission 
components can be summarized as follows.  Naively, Eq.(\ref{eq:generic_csection})
is finite for these contributions, and the limit $\epsilon \to 0$ can be taken.  However, 
this is only true for non-exceptional momentum configurations. If the momenta 
of some particles become soft, $q_i \to 0$, or collinear,
$q_i \cdot q_j \to 0,~q_{i,j} \ne 0$, the matrix element diverges
 and can not be integrated in four dimensions. Computing  
the contribution of the real emission graphs to the total cross section 
amounts to integrating  Eq.(\ref{eq:generic_csection}) over the entire phase space 
($F_j \to 1$), so 
that the soft and collinear regions are included. However, 
for the differential cross section such an integration is not allowed, 
since we want to keep the kinematics of the final state intact. 
The challenge is then to extract the singularities from Eq.(\ref{eq:generic_csection}) 
in a way that correctly accommodates both singular and non-singular 
limits, and does not require any integrations to be performed.
This can be done using the approach suggested in \cite{sector}, which we explain 
in detail in this paper.  We sketch here its outcome.   
Using this method, we are able to rewrite  ${\rm d}\sigma_{\rm real}$ 
in the following form:
\be
{\rm d}\sigma_{\rm real} = \sum \limits_{i=4}^{0} 
\frac{A_{i}[\{q_l\},F_{j}]}{\ep^{i}},
\label{eq4}
\ee
where $A_i$ are functions non-singular everywhere 
in the phase space.  No specific information about 
the measurement function $F_J$ is used in this derivation.
The functions $A_i$ contain no residual $\ep$-dependence, and 
can therefore be computed numerically in four dimensions.  
The expression for the double real emission component of Eq.(\ref{eq4})
is then combined with similar 
expressions for the virtual corrections, and the singularities 
in $\ep$ are canceled numerically.  After the cancellation 
of singularities is established, we can drop the singular 
terms from the expression for the cross section and implement 
the finite part into a numerical code.  This is the basic 
strategy which we discuss in detail in the remainder of this 
paper.


\section{Production of the Higgs boson in association with up 
to one parton}
\label{sec:soloHiggs}

We start with the partonic cross sections for producing
the Higgs boson and no partons in the final state:
\begin{equation}
g(p_1) + g(p_2) \to H(p_h).
\end{equation} 
The $2 \to 1$ phase-space is simple, because of momentum conservation. 
We derive
\begin{equation}
\int d\Pi_0 = \int d^dp_h \delta^d(p_h-p_1-p_2) \delta(p_h^2 - m_h^2)
=\delta(m_h^2-s),
\end{equation}
where $s=(p_1+p_2)^2$ 
is the partonic center of mass energy squared.

\begin{figure}
\begin{center}
\begin{picture}(300,100)(0,0)
\SetColor{Blue}
\SetWidth{1.0}

\Gluon(20,75)(50,50){2}{5}
\Gluon(20,25)(50,50){2}{5}
\put(47,47){$\otimes$}

\Gluon(110,80)(150,50){2}{6}
\Gluon(110,20)(150,50){2}{6}
\put(147,47){$\otimes$}
\Gluon(130,65)(130,35){2}{4}

\Gluon(210,80)(250,50){2}{6}
\Gluon(210,20)(250,50){2}{6}
\put(247,47){$\otimes$}
\Gluon(220,70)(220,30){2}{5}
\Gluon(230,65)(230,35){2}{4}

\SetColor{Red}
\SetWidth{1.5}
\Line(54,50)(75, 50)

\Line(154,50)(175, 50)

\Line(254,50)(275, 50)

\SetColor{Red}
\end{picture}
\end{center}
\caption{\label{fig:virtualdiagrams}Examples of  diagrams that contribute 
to the $g + g \to H$ cross section}
\end{figure}
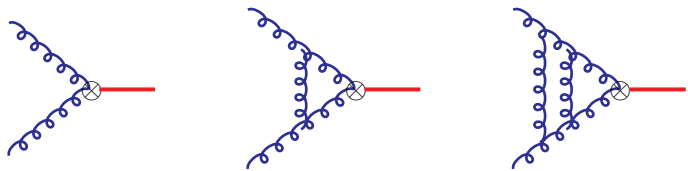

The most complicated part in computing 
the partonic channel $\sigma_{gg\to H}$ 
is the evaluation of the virtual 
corrections through  two-loops (see Fig.~\ref{fig:virtualdiagrams}). 
Fortunately, these corrections are known 
from the analytic calculation of the inclusive Higgs boson production 
cross section through NNLO
\cite{Harlander:2000mg,Harlander:2002wh,Anastasiou:2002yz,Ravindran:2003um},
and we use these results in this paper.

\begin{figure}
\begin{center}
\begin{picture}(300,100)(0,0)
\SetColor{Blue}
\SetWidth{1.0}

\Gluon(20,25)(49,25){3}{4}
\Gluon(20,75)(50,75){3}{4}
\Gluon(50,75)(80,75){3}{4}
\Gluon(50,75)(50,28){3}{6}
\put(47,23){$\otimes$}
\SetColor{Red}
\SetWidth{1.5}
\Line(56,25)(80, 25)

\SetColor{Blue}
\SetWidth{1.0}
\Gluon(200,25)(249,25){3}{7}
\Gluon(200,75)(250,75){3}{7}
\Gluon(220,25)(220,75){3}{6}
\Gluon(250,75)(280,75){3}{4}
\Gluon(250,75)(250,28){3}{6}
\put(247,23){$\otimes$}
\SetColor{Red}
\SetWidth{1.5}
\Line(256,25)(280, 25)
\end{picture}
\end{center}
\caption{\label{fig:realdiagrams}
Examples of diagrams that contribute to the production of the Higgs 
boson in association with one parton.}
\end{figure}
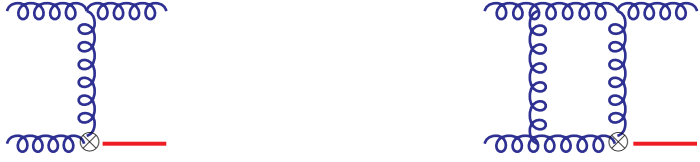

We next study the cross sections 
for partonic processes with the  Higgs boson and a quark or a gluon in 
the final state: $gg  \to Hg, qg \to Hq$, and $ q \bar{q} \to Hg$. For these 
processes, we must compute the corresponding tree-level and one-loop 
amplitudes. We 
consider the process $g(p_1) + g(p_2) \to H(p_h) + g(p_3)$ as an 
example. Typical diagrams are shown in Fig~\ref{fig:realdiagrams}.

Consider a contribution arising from the interference of two tree-level diagrams 
to the differential cross section. It can be written as:
\begin{equation}
\frac{{\cal N}(s_{13}, s_{23},F_J)}{s_{13} s_{23}}, 
\label{eq:ds}
\end{equation}
where $s_{ij} = (p_i-p_j)^2$ and $F_J$ is the measurement function which 
defines the observable we want to compute. The only information we 
need about the numerator in Eq.(\ref{eq:ds}) is that it is a finite 
function in the limits $s_{13} \to 0$ and $s_{23} \to 0$.
The structure  of the infrared and collinear 
singularities is fully determined 
by the denominator of Eq.(\ref{eq:ds}). We use this 
observation to derive an expansion 
in $\epsilon$ for this denominator in terms of delta functions 
and plus distributions. Having done that, we treat 
arbitrary numerators using a numerical subroutine 
for $F_J$, and defining procedures to compute the action of delta functions and 
plus distributions on integrable functions.

Therefore, the basic integral we have to consider is
\begin{eqnarray}
\label{eq:equatA}
 I_{gg \rightarrow Hg}=\int d^dp_h d^dp_3 \delta^+\left(p_h^2-m_h^2\right)
\delta^+\left(p_3^2\right) \delta^d\left(p_1+p_2-p_h-p_3\right) 
\frac{F_J(s_{13},s_{23})}{s_{13} s_{23}}.
\end{eqnarray}
This integral  is potentially singular for $s_{13},s_{23} = 0$.
To extract the singularities, we parameterize the phase-space in terms of 
variables that range from 0 to 1 in such a way that 
the singularities are mapped to 
the boundaries of the integration region. 
A convenient 
parameterization is in terms of the variables $\lambda_1, \lambda_2$, where 
\begin{equation}
s_{13} = -\frac{m_h^2}{\lambda_1} \frac{1+\lambda_1}{\lambda_1+\lambda_2}
\left(1-\lambda_1 \right),
~~~~s_{23} = -\frac{m_h^2}{\lambda_2} \frac{1+\lambda_2}{\lambda_1+\lambda_2}
\left(1-\lambda_2 \right).
\end{equation}
In this parameterization the integral in Eq.(\ref{eq:equatA})
becomes
\begin{eqnarray}
\label{eq:sigmaR}
I_{gg \rightarrow Hg} &=&  \frac{\Omega_{d-2}}{2 s} 
\int_0^1 d\lambda_1 d\lambda_2 
\delta\left(\lambda_1 \lambda_2 - \frac{m_h^2}{s}\right)
(1-\lambda_1)^{-1-\epsilon} (1-\lambda_2)^{-1-\epsilon} \nonumber \\
&& \frac{1+\lambda_1 \lambda_2}{(1+\lambda_1)(1+\lambda_2)}
\left[ 
\frac{m_h^4 (1+\lambda_1)(1+\lambda_2)}
{\lambda_1 \lambda_2 (\lambda_1 + \lambda_2)^2}
\right]^{-\epsilon} F_J(s_{13}, s_{23}). 
\end{eqnarray}
The delta function appears because of the 
 momentum conservation 
$s_{12}+s_{13}+s_{23}=m_h^2$, and prevents $\lambda_{1,2}$
from reaching $0$.  The singularities that occur as $\lambda_1,\lambda_2 \to 1$ are in a factorized form. 
To extract them, we rewrite the singular terms 
$(1-\lambda_1)^{-1-\epsilon}$, $(1-\lambda_2)^{-1-\epsilon}$ using 
\begin{equation}
\label{eq:plusexpansion}
\lambda^{-1+\epsilon} = \frac{\delta(\lambda)}{\epsilon} + 
\sum_{n=0}^{\infty} \frac{\epsilon^n}{n!} 
\left[\frac{\log(\lambda)^n}{\lambda} \right]_+,
\end{equation}
and  expand in $\epsilon$.
The result contains
delta functions 
and plus distributions and can be integrated numerically with the functions $F_J$. 

Finally, we must discuss the computation of the interference terms 
of the one-loop and tree-level amplitudes for Higgs boson production in 
association with a single parton. 
These terms require the calculation of the one-loop amplitude, and 
integrations over the $2\to 2$ phase-space variables. Using standard 
reduction methods, we write the one-loop amplitude in terms of 
master integrals that are known analytically. 
The integration over the phase-space then proceeds in a way described above.

\section{Computation of the collinear subtraction terms}
\label{sect:collinear}

In this Section we discuss the computation of the collinear 
subtraction terms.  We remind the reader that, by comparing 
the hadroproduction cross section written through ``bare'' and 
renormalized quantities,  a relation between 
the bare ($\sigma$)  and the renormalized ($\hat{\sigma}$)
partonic cross sections can be derived, as shown in Eq.(\ref{eq:sigma_convol}).
Having computed the bare cross section, we then derive the 
renormalized partonic cross section $\hat{\sigma}_{ij}$
by solving Eq.(\ref{eq:sigma_convol})
order-by-order in the expansion 
in $\alpha_s$.  We must consider the convolution integrals of the 
partonic cross sections with the Altarelli-Parisi kernels at each order in
the perturbative expansion. Technical aspects of this computation are 
discussed in this Section.

Through NNLO, we have to consider 
two distinct cases when $\hat \sigma_{ij}$ is either the 
leading order or the next-to-leading order cross section
discussed in the previous Section.
Consider first the leading order cross section. Since
$\sigma_{gg}^{(0)} \sim \delta(1-z)$, where $z = m_h^2/s$, 
the required convolution integrals are of the form:
\begin{eqnarray}
\label{eq:typical_convol}
I = \int_z^1 dy_1 \int_{\frac{z}{y_1}}^1 dy_2 \Gamma_a(y_1)  \Gamma_b(y_2) \delta(1-\frac{z}{y_1 y_2})  
=  z \left( \Gamma_a \otimes \Gamma_b \right)(z).
\end{eqnarray}
We therefore need to consider convolutions of the  splitting functions 
which appear in the perturbative expansion of $\Gamma_a,\Gamma_b$. 
The splitting functions generically contain delta-functions, 
plus-distributions and regular functions. The convolutions that 
involve delta-functions are straightforward. 
The convolutions 
of two regular functions and the convolutions of a plus-distribution 
with a regular function are performed numerically.
The convolution of two plus-distributions requires  additional 
analytic work. 

Consider the convolution of two plus-distributions:
\be
I_{nm} = \left[\frac{\ln^n(1-x)}{1-x} \right]_+ \otimes
\left[\frac{\ln^m(1-x)}{1-x} \right]_+.
\ee
We first apply the definition of plus-distributions in the
convolution integral:
\begin{eqnarray}
&& I_{mn} 
=  \int_0^1 dy dz \frac{\ln^n(1-y)}{1-y}
\frac{\ln^m(1-z)}{1-z}
\left\{\delta(x-yz)-\delta(x-y)-\delta(x-z)+\delta(x-1)\right\}.
\label{eq.aux}
\end{eqnarray}
Eq.(\ref{eq.aux}) can not be integrated term by term because 
of divergences; to make such an integration possible, we introduce auxiliary 
regularizations:
\begin{eqnarray}
\label{eq:regplusplus}
I_{nm} &=& \lim_{\epsilon \to 0} \lim_{{a,b \to 1}} 
 \frac{\partial^n}{\partial^n a} 
\frac{\partial^m}{\partial^m b} \frac{1}{\epsilon^{n+m}}
\int_0^1 dy dz (1-y)^{-1+a\epsilon} (1-z)^{-1+b\epsilon} \nonumber \\
&& \left\{\delta(x-yz)-\delta(x-y)-\delta(x-z)+\delta(x-1)\right\}.
\end{eqnarray}
The four integrals in Eq.(\ref{eq:regplusplus})
can now be computed separately. The first term
\begin{equation}
I_a = \int_0^1 dy dz (1-y)^{-1+a\epsilon} (1-z)^{-1+b\epsilon} 
\delta(x-zy),
\end{equation}
after integrating over $y$ and performing the change of variables
$z = x +(1-x) \lambda$, takes the form 
\begin{equation}
\label{eq:ref20}
I_a = (1-x)^{-1+(a+b)\epsilon} \int_0^1 d\lambda 
\left[x +(1-x) \lambda\right]^{-a\epsilon}
\lambda^{-1+a\epsilon} (1-\lambda)^{-1+b\epsilon}.
\end{equation}
We then rewrite Eq.(\ref{eq:ref20}) using Eq.(\ref{eq:plusexpansion}).  
The remaining three terms in Eq.(\ref{eq:regplusplus}) are easy to compute.
After differentiating the result with respect to $a,b$ and 
taking the limit $a,b \to 1$, we expand in $\epsilon$. The leading 
term in the $\epsilon$ expansion yields the convolution of the required 
plus distributions.  This solves the problem of computing the collinear 
factorization terms when the leading order cross section $\sigma_{gg}^{(0)} 
\sim \delta(1-z)$ is involved.

We now proceed to the discussion of the integrals that involve the 
NLO cross sections for the production of the Higgs boson in association 
with one additional parton.
Inserting Eq.(\ref{eq:sigmaR}) 
into Eq.(\ref{eq:sigma_convol}) we produce integrals of the form 
\begin{eqnarray}
\label{eq:lambda_convol}
\int_0^1 d\lambda_1 d\lambda_2  
dy_1 dy_2 \Gamma_1(y_1)
\Gamma_2(y_2)f(\lambda_1 \lambda_2)
\delta\left(y_1 y_2 \lambda_1 \lambda_2 - \frac{m_h^2}{s}\right),
\end{eqnarray} 
where $f(\lambda_1, \lambda_2)$ is the integrand of Eq.(\ref{eq:sigmaR}) 
without the delta function and  after the expansion  in $\epsilon$. 
The parameterization of the phase-space in terms of the variables 
$\lambda_1,\lambda_2$ is convenient for  rewriting the  
integral Eq.(\ref{eq:lambda_convol}) through 
successive conventional convolutions, 
\begin{eqnarray}
 \int \limits_{0}^{1} d\lambda_1 d\lambda_2
\delta\left(\lambda_1 \lambda_2 - \frac{m_h^2}{s}\right) 
\int \limits_{0}^{1} 
d\lambda_3 dy_1 \delta(\lambda_1 - y_1 \lambda_3)\Gamma_1(y_1)
\int \limits_{0}^{1} 
d\lambda_4 dy_1 \delta(\lambda_2 - y_2 \lambda_4)\Gamma_1(y_2)
f(\lambda_3, \lambda_4).
\nonumber 
\end{eqnarray}
We  note that since we work through the relative order 
${\cal O}(\alpha_s^2)$,  one of the two 
kernels $\Gamma(y_{1,2})$ in 
Eq.(\ref{eq:lambda_convol})
 is a delta function, so that 
the corresponding integration can easily be performed. After 
that, we are left with a convolution integral that 
can be treated along the lines described
at the beginning of this Section.  It can then be directly used in the 
numerical integration with 
parton distribution functions.

 \section{Phase space parameterizations for double real emission processes}
\label{sect:phspacepar}

We can now study the contributions to the NNLO cross section from processes 
with double real emissions. These processes appear only at tree-level.  However, the 
integrations over the $2 \to 3$ phase-space 
are involved.  Our aim in the following sections is to present a 
detailed description of their treatment.

The first step in using the method described in \cite{sector} is to choose 
a parameterization of the double real emission phase space in which the 
integration region is the unit hypercube.  This is required in order to 
use an expansion in plus distributions to extract singularities.      
In principle, any parameterization 
that accomplishes this is acceptable.  
In practice, 
finding a convenient one that reduces the number of sector 
decompositions is important for the efficiency of the approach.  We will 
discuss here how to choose a parameterization suitable for the topologies 
which contribute to the Higgs production process. 

For the double real emission corrections to Higgs production at NNLO, 
we must parameterize a $2 \rightarrow 3$ particle phase space, with 
one massive final-state particle.  We consider here 
$g(p_1) + g(p_2) \to H(p_h) + g(p_3) + g(p_4)$ as a prototypical partonic 
process, although the formulae we derive are valid for all such partonic 
processes.  We consider a fixed energy for the partonic collision, 
$(p_1+p_2)^2 = s$.  
The scalar products that appear in the matrix elements are 
\begin{itemize}
 \item $s_{if} = (p_i-p_f)^2$, where $i=1,2$ and $f=3,4$;
 \item $s_{34} = (p_3+p_4)^2$;
 \item $s_{ih} = (p_i-p_h)^2$, where $i=1,2$;
 \item $s_{hf} = (p_f+p_h)^2$, where $f=3,4$.
\end{itemize}
The invariant masses of the form $s_{hf}$ are bounded 
from below by $s_{hf} \geq m_{h}^2$, and therefore do not lead to singularities.  
The $2 \rightarrow 3$ phase space which we must map to the unit hypercube 
is 
\begin{equation}
{\rm d\Pi} = \int [dp_3][dp_4][dp_h] \delta^{(d)}(p_1+p_2 - p_h - p_3 - p_4).
\label{phasestart}
\end{equation}
where $[dq] = {\rm d}q^{d-1}/(2q_0)$. 
We will set the overall scale $s=1$ in the discussions that follow; it can 
be restored by dimensional arguments.  We also denote a generic 
invariant mass by $s_{ab}$; the subscripts 
$i,f,h$ will be used as in the above list.  Therefore, $i=1,2$ and $f=3,4$.

We denote the variables that describe the unit hypercube by $\lambda_i$; the partonic 
phase space is spanned by four independent variables, so $i=1,\ldots,4$.  
The singularity 
structure is dictated by the invariant masses that appear in the denominator of the 
matrix elements. In terms of these variables, the invariant masses can take one of the following three 
generic forms, ordered from most to least desirable:
\begin{itemize}
 \item a factorized form, in which 
  \begin{equation}
    \frac{1}{s_{ab}} = \frac{1}{\lambda_1\lambda_2 \ldots};
  \end{equation}
 \item an entangled form, in which
   \begin{equation}
    \frac{1}{s_{ab}} = \frac{1}{(\lambda_1+\lambda_2) \ldots};
\label{entg}
  \end{equation}
 \item a ``line'' singularity, in which
   \begin{equation}
    \frac{1}{s_{ab}} = \frac{1}{|\lambda_1-\lambda_2| \ldots}.
   \end{equation}
\end{itemize}
The singularities as $\lambda_1,\lambda_2 \rightarrow 0$ in the factorized form  can 
be immediately extracted using the expansion in plus distributions of Eq.(\ref{eq:plusexpansion}).
We will discuss the remaining two structures in detail later in the text, and 
we only briefly describe here how they are dealt with. 
In order to extract the singularity from the entangled form in Eq.(\ref{entg}), we must 
sector decompose in the variables $\lambda_1$ and $\lambda_2$; 
this involves splitting the 
integration region into two sectors, 
$\lambda_1 > \lambda_2$ and $\lambda_2 > \lambda_1$, 
remapping the integration limits to $[0,1]$, and then expanding the resulting 
expressions in plus distributions.  For the line singularity, in which the singular region 
is along an entire edge of phase space rather than just a point, we must first perform an 
additional variable change to remap the singular region to a point, and then typically 
perform a series of sector decompositions.
It is advantageous to express as many of the singular structures in a factorized form as possible; this form gives only 
one sector rather than several, which leads to smaller and typically simpler 
expressions.  It also preserves the kinematics of the parameterization, {\it i.e.}, 
there is only one mapping between the $s_{ij}$ and the $\lambda_i$.  In the 
other forms, each sector has a different kinematics.

In order to choose a convenient set of parameterizations, 
we must first study the 
double real emission diagrams 
that contribute to Higgs production.  
There are three generic 
types that must be considered:
\begin{enumerate}
 \item Diagrams ($C_1$) in which the Higgs is emitted from either an internal or final-state line, and the 
   singular invariant masses that can appear in the denominator are $s_{if}$;
     \begin{center}
      \begin{picture}(50,50)(0,0)
        \SetColor{Blue}
        \Gluon(0,50)(25,50){3}{6}
        \Gluon(25,50)(50,50){3}{6}
       \Gluon(0,0)(25,0){3}{6}
       \Gluon(25,0)(50,0){3}{6}
       \Gluon(25,0)(25,22){3}{6}
       \Gluon(25,28)(25,50){3}{6}
        \put(21,23){$\otimes$}
        \SetWidth{1.5}
        \SetColor{Red}
        \Line(28,25)(50,25)
        \put(-10,-2){2}
        \put(-10,48){1}
        \put(58,-2){4}
        \put(58,48){3}
        \put(58,23){$h$}
      \end{picture}
\hspace{2cm}
      \begin{picture}(50,50)(0,0)
        \SetColor{Blue}
        \Gluon(0,50)(15,50){3}{2}
        \Gluon(15,50)(30,50){3}{4}
        \Gluon(35,50)(50,50){3}{4}
       \Gluon(0,0)(15,0){3}{2}
       \Gluon(15,0)(50,0){3}{8}
       \Gluon(15,0)(15,50){3}{12}
        \put(30,48){$\otimes$}
        \SetWidth{1.5}
        \SetColor{Red}
        \Line(34,46)(50,25)
        \put(-10,-2){2}
        \put(-10,48){1}
        \put(58,-2){4}
        \put(58,48){3}
        \put(58,23){$h$}
      \end{picture}
     \end{center}
 \item Diagrams ($C_2$) for which the potential singular denominators are $s_{34}$ and $s_{ih}$;
     \begin{center}
      \begin{picture}(50,50)(0,0)
        \SetColor{Blue}
        \Gluon(0,50)(15,50){3}{2}
        \Gluon(15,50)(32.5,50){3}{4}
        \Gluon(32.5,50)(53,52){3}{3.5}
       \Gluon(0,0)(13,0){3}{2}
       \Gluon(32.5,50)(50,25){3}{6}
       \Gluon(15,2)(15,50){3}{12}
        \put(13,-2){$\otimes$}
        \SetWidth{1.5}
        \SetColor{Red}
        \Line(20,0)(50,0)
        \put(-10,-2){2}
        \put(-10,48){1}
        \put(58,-2){$h$}
        \put(58,48){3}
        \put(58,23){4}
      \end{picture}
     \end{center}
 \item Diagrams ($C_3$) in which the Higgs is emitted from an 
initial-state line, and the singular invariant masses that can appear 
in the denominator are $s_{if}$ and $s_{ih}$;
     \begin{center}
      \begin{picture}(50,50)(0,0)
        \SetColor{Blue}
        \Gluon(0,50)(25,50){3}{6}
        \Gluon(25,50)(50,50){3}{6}
       \Gluon(0,0)(23,0){3}{6}
       \Gluon(25,0)(25,25){3}{6}
       \Gluon(25,27)(25,50){3}{6}
       \Gluon(25,25)(50,25){3}{6}
        \put(23,-2){$\otimes$}
        \SetWidth{1.5}
        \SetColor{Red}
        \Line(30,0)(50,0)
        \put(-10,-2){2}
        \put(-10,48){1}
        \put(58,-2){$h$}
        \put(58,48){3}
        \put(58,23){4}
      \end{picture}
        \end{center}
\end{enumerate}

The matrix elements consist of interferences between these three classes of diagrams.
We can not find a phase-space parameterization in which all of the singular scalar products 
take a factorized form, so it is useful to consider parameterizations tailored to each 
structure.  We will discuss two different parameterizations of 
the partonic phase space, which we refer to as the ``energy'' and 
``rapidity'' parameterizations.

In the energy parameterization, the invariant masses $s_{if}$ take on simple, 
factorized forms; it is therefore useful for interferences within the first class of 
diagrams listed above.  It is named for the observation that in terms of energies 
and angles with respect to the initial beam axis, 
$s_{if} = -2E_iE_f[1 - {\rm cos}\theta_{if}]$; 
the soft and collinear singularities in these scalar products are therefore factorized.  To derive 
the expression for the phase space in the energy parameterization, we begin with Eq.(\ref{phasestart}), 
and use the momentum-conserving $\delta$-function to remove the $[dh]$ integration; we obtain
\begin{equation}
\label{pare0}
 d\Pi_{E} = \frac{1}{4} \int {\rm d}E_3 {\rm d}E_4 
d\Omega_3 d\Omega_4 \, E_{3}^{d-3} E_{4}^{d-3} 
 \delta\left[1-z+s_{13}+s_{14}+s_{23}+s_{24}+s_{34} \right],
\end{equation}
where $z=m_{h}^{2}/s$.  It is convenient to continue the calculation in the 
partonic center-of-mass (CM) frame.  We introduce 
the CM frame parameterization 
\begin{eqnarray}
p_1 &=& \frac{1}{2}\left(1,\vec{0},1\right),~~~~~
p_2 = \frac{1}{2}\left(1,\vec{0},-1\right), \nonumber \\ 
p_3 &=& E_{3}\left(1,{\rm sin}\theta_3,0,{\rm cos}\theta_3 \right),~~~
p_4 = E_{4}\left(1,{\rm sin}\theta_4{\rm cos}\phi,
{\rm sin}\theta_4{\rm sin}\phi,{\rm cos}\theta_4 \right).
\end{eqnarray}
In the expressions for $p_3$ and $p_4$ we have suppressed the additional $\epsilon$-dimensional components of 
the momenta; they can be chosen to vanish in this frame.  We use the $\delta$-function to remove the 
$E_4$ integration, which sets 
\begin{equation}
E_4 = \frac{1-z-2E_3}{2\left[1-E_{3}(1-\vec{n}_{1}\cdot \vec{n}_{2})\right]}
\label{E4expr}
\end{equation}
and gives the jacobian $2\left[1-E_{3}(1-\vec{n}_{1}\cdot \vec{n}_{2})\right]$;
in these expressions,
\begin{equation}
\vec{n}_{1}\cdot \vec{n}_{2} = {\rm cos}\theta_{3}{\rm cos}\theta_{4}
+{\rm sin}\theta_{3}{\rm sin}\theta_{4}{\rm cos}\phi .
\end{equation}

We must now consider the angular integrations $d\Omega_3$ and $d\Omega_4$.  In terms of the polar angles 
$\theta_3$ and $\theta_4$, and the azimuthal angle $\phi$, they become
\begin{eqnarray}
d\Omega_3 &=& {\rm d}{\rm cos}\theta_3 \left[{\rm sin}^{2}\theta_3 \right]^{-\epsilon} \Omega_{d-2} , \nonumber \\ 
d\Omega_4 &=& {\rm d}{\rm cos}\theta_4 {\rm d}{\rm cos}\phi  \left[{\rm sin}^{2}\theta_4\right]^{-\epsilon} 
 \left[{\rm sin}^{2}\phi\right]^{-\epsilon-1/2}\Omega_{d-3},
\end{eqnarray}
where $\Omega_d$ is the solid angle in $d$-dimensions:
\begin{equation}
\Omega_d = \frac{2\pi^{d/2}}{\Gamma(d/2)}.
\end{equation}
Using these expressions, we can write the phase space in the following form:
\begin{eqnarray}
d\Pi_{E} &=& \frac{\Omega_{d-2}\Omega_{d-3}}{8} 
\int {\rm d}E_3 {\rm d}{\rm cos}\theta_3
{\rm d}{\rm cos}\theta_4 {\rm d}{\rm cos}\phi \, 
  E_{3}^{d-3} E_{4}^{d-3} \left[{\rm sin}^{2}(\theta_3)\right]^{-\epsilon} \nonumber \\ & & \times 
\left[{\rm sin}^{2}\theta_4\right]^{-\epsilon} 
 \left[{\rm sin}^{2}\phi\right]^{-\epsilon-1/2} / \left[1-E_{3}(1-\vec{n}_{1}\cdot \vec{n}_{2})\right].
\end{eqnarray}
The angular integrations clearly range from $[-1,1]$.  To derive the limits of the $E_3$ integration, we note that 
$E_3,E_4 \geq 0$; using Eq.(\ref{E4expr}), we find that this implies
\begin{equation}
 0 \leq E_3 \leq \frac{1-z}{2} = E_{3}^{+}.
\end{equation}

We are now ready to map the integration region into the unit hypercube.  Performing the variable changes 
\begin{eqnarray}
E_3 &=& \lambda_1 E_{3}^{+},~~~~{\rm cos}\theta_3 = -1+2\lambda_2, \nonumber \\ 
{\rm cos}\theta_4 &=& -1+2\lambda_3,~~~~ {\rm cos}\phi = -1+2\lambda_4,
\end{eqnarray}
we derive the final expression for the partonic phase space in the energy parameterization:
\begin{eqnarray}
d\Pi_{E} &=& N \int \limits_{0}^{1} {\rm d}\lambda_1 {\rm d}\lambda_2 
{\rm d}\lambda_3 {\rm d}\lambda_4 [\lambda_1(1-\lambda_1)]^{1-2\ep}
[\lambda_2(1-\lambda_2)]^{-\ep} 
 [\lambda_3(1-\lambda_3)]^{-\ep} \nonumber \\
&& \times [\lambda_4(1-\lambda_4)]^{-\ep - 1/2}D^{2-d},
\label{enparam}
\end{eqnarray}
with 
\begin{eqnarray}
N &=& \Omega_{d-2} \Omega_{d-3} (1-z)^{3-4\epsilon}/2^{4+2\epsilon}, \nonumber \\  
D &=& 1 - (1-z)\lambda_1\left(1-\vec{n}_{1}\cdot \vec{n}_{2}\right)/2, \nonumber \\
1-\vec{n}_{1}\cdot \vec{n}_{2} &=& 2 \left[\lambda_2 + \lambda_3  
-2\lambda_2 \lambda_3 + 2 (1-2\lambda_4) \sqrt{\lambda_2(1-\lambda_2)\lambda_3
(1-\lambda_3)} \right].
\label{n1n2expr}
\end{eqnarray}
We note that the factor $D$ which appears in the phase space is bounded from 
below by $D \geq z$, and does not contribute to the singularity structure.  All limits 
$\lambda_i \rightarrow 0,1$ are separated and regulated by $\ep$, as required for the 
extraction of singularities shown in Eq.(\ref{eq:plusexpansion}).

We now discuss the singularity structure of the scalar products in this paramterization.  
As desired, the invariant masses $s_{if}$ take on simple, factorized forms:
\ba
&& s_{13} = -(1-z)\lambda_1(1-\lambda_2),~~~s_{23} = -(1-z)\lambda_1 \lambda_2,
\nonumber \\
&& s_{14} = -(1-z)(1-\lambda_1)(1-\lambda_3)/D,~~~s_{24} = -(1-z)(1-\lambda_1)\lambda_3/D.
\label{eninvm1}
\ea
The other invariant masses have more complex singular structures.  We find 
\ba
s_{34} &=& (1-z)^2\lambda_1(1-\lambda_1)\left(1-\vec{n}_{1}\cdot \vec{n}_{2}\right)/2/D, 
\nonumber \\  s_{1h} &=& -(1-z)\left\{\lambda_1\lambda_2+\lambda_3(1-\lambda_1)-(1-z)\lambda_1
\left[1-\lambda_1(1-\lambda_2)\right]\left(1-\vec{n}_{1}\cdot \vec{n}_{2}\right)/2\right\}/D,
\nonumber \\  s_{2h} &=& -(1-z)\left\{\lambda_1(1-\lambda_2)+(1-\lambda_3)(1-\lambda_1)-(1-z)\lambda_1
\left[1-\lambda_1\lambda_2\right]\left(1-\vec{n}_{1}\cdot \vec{n}_{2}\right)/2\right\}/D.\;\;\;\;\;\;\;
\label{eninvm2}
\ea
The $(1-\vec{n}_{1}\cdot \vec{n}_{2})$ factor in the numerator of $s_{34}$ leads to a line singularity 
along the edge of phase space where $\lambda_4=1$ and $\lambda_2=\lambda_3$; setting $\lambda_4=1$ 
in Eq.(\ref{n1n2expr}), we can write
\begin{equation}
\left(1-\vec{n}_{1}\cdot \vec{n}_{2}\right)|_{\lambda_4=1} = \left[\sqrt{\lambda_2(1-\lambda_3)}
 -\sqrt{\lambda_3(1-\lambda_2)}\right]^{2}.
\end{equation}
The invariant masses $s_{1h}$ and $s_{2h}$ do not contain line singularities, 
but do contain several entangled singularities; for example, $s_{1h}$ vanishes at the following 
phase-space boundaries: (1) $\lambda_1=1$ and $\lambda_2=0$; (2) $\lambda_1=0$ and $\lambda_3=0$;    
(3) $\lambda_2=0$ and $\lambda_3=0$.  These problems indicate that the energy parameterization is 
not well suited for interferences between diagrams of the second and third classes listed above.

For interferences between diagrams in the second and third classes, it is usually better to use the 
rapidity parameterization.  With this choice, the invariant masses $s_{34}$, $s_{ih}$, and two of 
the $s_{if}$ take on simple, factorized forms.  This parameterization is the fully differential 
extension of the one used in \cite{Anastasiou:2003ds} to calculate the NNLO corrections to Drell-Yan 
production of lepton pairs.

We begin by making explicit what variables we use to describe the phase space.  The four independent 
variables we choose are the invariant masses $s_{34}$, $s_{13}$, $s_{23}$, and the variable 
$u = p_{1}\cdot p_h /p_{2}\cdot p_h$.  Anticipating the translation of our partonic expressions 
into hadronic results, we note that $u$ is related to the lab-frame 
rapidity of the Higgs boson
by $u=x_{1} e^{-2Y}/x_{2}$, where $Y$ is the rapidity and $x_1$ and $x_2$ are the standard 
Bjorken $x$ variables for each initial-state parton.  We rewrite the phase space in Eq.(\ref{phasestart}) 
as 
\begin{eqnarray}
{\rm d\Pi_R} &=& \int {\rm d}s_{34}\,{\rm d}u\,{\rm d}s_{13}\,{\rm d}s_{23}\int [dp_3][dp_4][dh] 
\delta(s_{34}-2p_3\cdot p_4)\,\delta(u-p_1\cdot p_h/p_2\cdot p_h) \nonumber \\ 
& & \times \delta(s_{13}+2p_1\cdot p_3)\,\delta(s_{23}+2p_2\cdot p_3)\,\delta^{(d)}(p_1+p_2 - p_h - p_3 - p_4).
\label{rapstart}
\end{eqnarray}
It is useful to view the production of the final-state particles $p_3$, $p_4$, and $p_h$ as an iterative process; 
first, $p_h$ and the massive ``particle'' $Q_{34}=p_3+p_4$ are produced; then, $Q_{34}$ decays into 
$p_3$ and $p_4$.  This motivates the following nested decomposition of the phase space:
\begin{equation}
{\rm d\Pi_R} = \int ds_{34}\,du\,ds_{13}\,ds_{23} \,{\rm d\Pi_1}{\rm d\Pi_2},
\label{psdecomp}
\end{equation}
with 
\begin{eqnarray}
{\rm d\Pi_1} &=& \int [dh]{\rm d}^{d}Q_{34} \delta(Q^{2}_{34}-s_{34}) \, \delta(u-p_1\cdot p_h/p_2\cdot p_h) \,
\delta^{(d)}(p_1+p_2-p_h-Q_{34}), \nonumber \\ 
{\rm d\Pi_2} &=& \int [dp_3][dp_4] \delta(s_{13}+2p_1\cdot p_3)\,\delta(s_{23}+2p_2\cdot p_3)\, 
\delta^{(d)}(Q_{34}-p_3-p_4).
\end{eqnarray}

We evaluate ${\rm d\Pi_1}$ in the CM frame of the $p_1+p_2$ system.  The $\delta$-functions remove all integrations 
except for those describing the azimuthal angles, which give only an overall solid angle factor.  We arrive at
\begin{equation}
{\rm d\Pi_1}=\frac{\Omega_{d-2}(1+z-s{34})}{4(1+u)^2}
\left[\frac{us_{34}^2-2u(1+z)s_{34}+(u-z)(1-uz)}{(1+u)^2}\right]^{-\ep}.
\label{pi1expr1}
\end{equation}
The $\ep$-dependent factor will be needed to regulate singularities in $u$, $z$, and $s_{34}$, as shown in 
Eq.(\ref{eq:plusexpansion}); however, the singular structures of these variables are entangled, and must be separated.  
To do so, note that the bracketed term in Eq.(\ref{pi1expr1})
can be written as 
\begin{equation}
\label{hpt}
\frac{u(s_{34}^{+}-s_{34})(s_{34}^{-}-s_{34})}{(1+u)^2},
\end{equation}
with $s_{34}^{\pm}=(r\pm t)(1\pm rt)/r$, $r=\sqrt{u}$, and $t=\sqrt{z}$.  
Eq.(\ref{hpt}) is the $p_{\perp}^{2}$ of 
the Higgs, and must be positive definite.  We must therefore demand $0 \leq s_{34} \leq s_{34}^{-}$.  We set 
$s_{34} = \lambda_1 s_{34}^{-}$, where $\lambda_1$ is in the range [0,1], and derive 
\begin{equation}
{\rm d\Pi_1}=\frac{\Omega_{d-2}(1+z)}{4(1+u)^2}\left\{1-\frac{\lambda_1 K_m}{r(1+z)}\right\}
\left[\frac{(u-z)(1-uz)(1-\lambda_1)(1-\lambda_1 K_m/K_p)}{(1+u)^2}\right]^{-\ep},
\label{pi1expr2}
\end{equation}
where $K_{p,m}=(r \pm t)(1 \pm rt)$.  The factor $K_m/K_p <1$, and the expression in the curly 
brackets does not vanish, so the $\lambda_1 \rightarrow 1$ limit has been separated from $u,z$.  However, 
the limits $u \rightarrow z,1/z$ and $z \rightarrow 1$ have not been separated, as is clear from 
the $(u-z)$ and $(1-uz)$ factors in Eq.(\ref{pi1expr2}); these will appear in the denominator and 
lead to singularities, so they must be dealt with.  We first note that to keep $p_{\perp}^{2} \geq 0$ 
we must have $z \leq u \leq 1/z$. We then 
follow \cite{ellis,Anastasiou:2003ds} 
and set 
\begin{equation}
y = \frac{u-z}{(1-z)(1+u)},
\end{equation}
with $y$ in the range $[0,1]$.  The phase space becomes
\begin{equation}
{\rm d\Pi_1}=\frac{\Omega_{d-2}(1+z)}{4(1+u)^2}\left\{1-\frac{\lambda_1 K_m}{r(1+z)}\right\}
\left[y(1-y)(1-z)^2(1-\lambda_1)(1-\lambda_1 K_m/K_p)\right]^{-\ep};
\end{equation}
the limits $u \rightarrow z,1/z$ have been separated from $z \rightarrow 1$ and moved to $y \rightarrow 0,1$.

Having discussed ${\rm d\Pi_1}$, we now consider ${\rm d\Pi_2}$.
It is again convenient to work in the CM frame of the $p_1+p_2$ system. The 
momenta can be written as 
\begin{eqnarray}
p_1 &=& \frac{1}{2}\left(1,\vec{0},1\right),~~~~~p_2 = \frac{1}{2}\left(1,\vec{0},-1\right), \nonumber \\ 
p_3 &=& \left(E,p_{\perp}{\rm sin}\theta,p_{\perp}{\rm cos}\theta,p_z\right),
\end{eqnarray}
where we have again suppressed the $\epsilon$-dimensional components of the momenta.  The $\delta$-functions 
present in ${\rm d\Pi_2}$ constrain $E$, $p_z$, $p_{\perp}$, and $\theta$.  Using these to remove the 
integrations, we arrive at
\begin{equation}
{\rm d\Pi_2} = \frac{\Omega_{d-3}(1+u)}{8\sqrt{u(s^{+}_{34}-s_{34})(s^{-}_{34}-s_{34})}} 
\left[p_{\perp}^{2}{\rm sin}^{2}\theta\right]^{-\ep-1/2}.
\end{equation}
To express $p_{\perp}{\rm sin}\theta$ through the variables $s_{34}$, $s_{23}$, $s_{13}$, and $u$, we 
must first define the auxiliary variables
\begin{eqnarray}
s_{13}^{\pm} &=& \frac{s_{34}(1+u)[s_{23}(1+u)+s_{34}+u-z]-us_{23}(s^{+}_{34}-s_{34})(s^{-}_{34}-s_{34}) \pm \sqrt{D_2}}
{(s^{-}_{23})^2(1+u)^2}, \nonumber \\ 
D_2 &=& -16u(1+u)^2s_{23}s_{34}(s^{+}_{34}-s_{34})(s^{-}_{34}-s_{34})(s_{23}^{-}+s_{23}), \nonumber \\ 
s^{-}_{23} &=& y(1-z)\left\{1+\frac{\lambda_1(1-rt)}{r(r+t)}\right\}.
\label{auxexpr}
\end{eqnarray}
In terms of these variables, we have 
\begin{equation}
p_{\perp}^{2}{\rm sin}^{2}\theta = \frac{-(s^{-}_{23})^2(1+u)^2(s^{+}_{13}+s_{13})(s^{-}_{13}+s_{13})}
{4u(s^{+}_{34}-s_{34})(s^{-}_{34}-s_{34})}.
\label{ptsexpr}
\end{equation}
There are several consistency conditions that we must apply to these expressions; these will give us 
the integration limits for $s_{13}$ and $s_{23}$.  First, to ensure the reality of $s^{\pm}_{13}$ in 
Eq.(\ref{auxexpr}), we must demand $D_2 \geq 0$; this implies $0 \geq s_{23} \geq -s^{-}_{23}$.  Finally, 
since $p^{2}_{\perp}{\rm sin}^{2}(\theta) \geq 0$, it is clear from Eq.(\ref{ptsexpr}) that we must 
require $-s^{+}_{13} \geq s_{13} \geq -s^{-}_{13}$.  With these limits, we can map the $s_{23}$ and $s_{13}$ 
integrations to the region [0,1].  We set 
\begin{equation}
s_{23} = -\lambda_2 s^{-}_{23},~~~~ s_{13}=-\lambda_4(s^{+}_{13}-s^{-}_{13})-s^{-}_{13},
\end{equation}
and derive the final expression for ${\rm d\Pi_2}$ in terms of $\lambda_1$, $\lambda_2$, $\lambda_4$, and $y$:
\begin{equation}
{\rm d\Pi_2} = \frac{\Omega_{d-3}2^{-4-2\ep}}{(1-z)\sqrt{y(1-y)(1-\lambda_1)(1-\lambda_1K_m/K_p)}}
\left[\frac{y(1-y)(1-z)^2\lambda_1\lambda_2(1-\lambda_2)\lambda_4(1-\lambda_4)}{r(r+t)(1+rt)}\right]^{-\ep-1/2}.
\end{equation}

Finally, we must combine ${\rm d\Pi_1}$ and ${\rm d\Pi_2}$ as shown in Eq.(\ref{psdecomp}), and include the 
jacobian for the transformation $\int ds_{34}\,du\,ds_{13}\,ds_{23} \rightarrow \int d
\lambda_1\,d\lambda_2\,d\lambda_4\,dy$.
This is simply done using the variable changes given above, and we derive the following final expression for the 
phase space in the rapidity parameterization:
\ba
&& {\rm d\Pi_R} = N \int \limits_{0}^{1} {\rm d}\lambda_1 {\rm d}\lambda_2 
{\rm d}\lambda_3 {\rm d}\lambda_4 \left[(1-\lambda_1)(1-\lambda_1 K_m/K_p)\right]^{-\epsilon} 
\nonumber \left[\lambda_1\lambda_2(1-\lambda_2)\right]^{-\epsilon} \nonumber \\ & & \times
\left[\lambda_3 (1-\lambda_3)\right]^{1-2\epsilon}
\left[\lambda_4(1-\lambda_4)\right]^{-\epsilon-1/2} 
\left[K_p r/(1+u)^2\right]^{-1+\epsilon} \left[1-\frac{\lambda_1 K_m}{r(1+z)}\right].
\label{rapparam}
\ea
We have changed $y \rightarrow \lambda_3$ in this equation for consistency of 
notation.  All the limits $\lambda_i \rightarrow 0,1$ have been separated.  
The singular scalar products in this parameterization are
\ba 
s_{1h} &=& -\lambda_3 (1-z)\left[1-\lambda_1 r(1-rt)/(r+t)\right], \nonumber \\ 
s_{2h} &=& -(1-\lambda_3) (1-z)\left[1-\lambda_1 (r-t)/r/(1+rt)\right], \nonumber \\ 
s_{23} &=& -\lambda_2 \lambda_3 (1-z) \left[1+\lambda_1 (1-rt)/r/(r+t)\right], \nonumber \\
s_{24} &=& -(1-\lambda_2) \lambda_3 (1-z) \left[1+\lambda_1 (1-rt)/r/(r+t)\right], \nonumber \\
s_{34} &=& \lambda_1 \lambda_3 (1-\lambda_3) (1-z)^2 (1+u)^2/K_p/r, \nonumber \\ 
s_{13} &=& -\frac{(1-\lambda_3)(1-z)}
{K_p r \left[1+\lambda_1 (1-rt)/r/(r+t)\right]} \left[A_1+A_2+2(2\lambda_4-1)\sqrt{A_1A_2}\right],
\label{rapinvm}
\ea 
where 
\begin{equation}
A_1 = \lambda_1(1-\lambda_2)(1+u)^2,~~~~A_2 = \lambda_2(1-\lambda_1)r(K_p-\lambda_1 K_m).
\label{Adef}
\end{equation}
We have not included $s_{14}$ in this list; it can be derived from the 
other invariant masses (see Eq.(\ref{pare0})), 
and we will see later that the parameterization can always be chosen so that it never 
appears in the denominator.  As claimed above, the invariant masses $s_{1h}$, $s_{2h}$, $s_{23}$, $s_{24}$, and $s_{34}$ are 
all in factorized forms; none of the expressions in square brackets for these invariant masses 
in Eq.(\ref{rapinvm}) vanish.  However, $s_{13}$ clearly has a line singularity when 
$\lambda_4=0$ and $A_1=A_2$; in terms of the $\lambda_i$, 
\begin{equation}
A_1=A_2 \Rightarrow \lambda^{s}_2 = \frac{\lambda_1(1+u)^2}{\lambda_1(1+u)^2+(1-\lambda_1)r(K_p-\lambda_1 K_m)} \le 1,
\label{Aphys}
\end{equation}
so this singularity occurs in the physical region.

Before concluding our presentation of the phase space parameterizations, and beginning our 
discussion of how to handle the entangled scalar products and line singularities, we must 
discuss what happens to the variable $z$ when we use our results to derive the hadronic cross section.  
When we convolute the partonic cross sections with the 
parton distribution functions to form the 
hadronic cross section, the variable $z$ scales as 
$z \rightarrow m_{h}^2/(x_1x_2s_{had})$, where 
$s_{had}$ is the hadronic center-of-mass energy squared, and the $x_i$ are the 
fractions of the hadronic momenta carried into the hard scattering process.  It is clear from 
the invariant masses in Eqs.(\ref{eninvm1},\ref{eninvm2},\ref{rapinvm}) that the matrix 
elements will contain singularities as $z \rightarrow 1$.  However, it can also be seen 
from these equations that this singularity is always in a factorized form.  The normalization factor 
$N$ in Eq.(\ref{n1n2expr}) 
contains the factor $(1-z)^{-4\epsilon}$, which regulates this limit for both 
parameterizations.  Singularities in $z$ and $\lambda_i$ can 
therefore be treated identically.

\section{Factorizing singularities}
\label{sect:factsing}

After choosing a phase-space parameterization for a given 
term from the matrix elements, we 
must extract its singularities without actually integrating over the $\lambda_i$.  We then have 
differential distributions in the $\lambda_i$\, giving us complete control over the kinematics and allowing 
us to compute arbitrary differential observables.  For factorized singularities, this is simple; the 
expansion in plus distributions shown in Eq.(\ref{eq:plusexpansion}) extracts the singularity as a $1/\ep$ pole 
multiplied by a $\delta$-function which restricts the integration to the singular region of phase space.  
The singularities can be cancelled numerically and then discarded, as shown in \cite{sector,Anastasiou:2004qd,higgsdiff}, 
leaving a finite, fully differential cross section.  Our goal will be to reduce all other singularities to 
a factorized form.

Before discussing the detailed procedure we use to handle entangled and line singularities, we must first 
explain sector decomposition \cite{Binoth:2000ps,hepp,Roth:1996pd}.  This is our primary technique of separating and 
factorizing entangled singularities.  We can illustrate the main features of sector decomposition with 
a simple example.  Consider the integral 
\begin{equation}
I = \int_{0}^{1} {\rm d}x{\rm d}y \frac{x^{\ep}y^{\ep}}{(x+y)^2} .
\label{eqexs}
\end{equation}
If we naively apply the expansion of Eq.(\ref{eq:plusexpansion}), we would conclude that we can simply Taylor-expand the 
numerator of the integrand, and would arrive at
\begin{equation}
I_{{\rm naive}} = \int_{0}^{1} {\rm d}x{\rm d}y \frac{1}{(x+y)^2}
\left\{1
+{\cal O}(\ep) \right\}.
\end{equation}
Integrating the ${\cal O}(\ep^0)$ term over $x$, we obtain
\begin{equation}
I_{{\rm naive}} = \int_{0}^{1}{\rm d}y \frac{1}{y(1+y)} + {\cal O}(\ep).
\end{equation}
This is divergent as $y \rightarrow0$; we have clearly missed 
a singularity. A simple analysis of Eq.(\ref{eqexs}) shows 
that the integral  diverges logarithmically  in the limit 
$x \sim y \rightarrow 0$;
if we attempt to study these limits separately, as we did 
in $I_{{\rm naive}}$, we miss this singularity.  
To deal with the singular phase-space region $x \sim y \sim 0$
we use sector decomposition.  

To introduce this technique, we will 
consider the same integral as in Eq.(\ref{eqexs}), 
but with an additional function in the integrand: 
$F_J[s_{ab}(x,y)]$, a measurement function, which 
describes kinematic features of the process such as dependence
on the parton distribution functions and phase-space constraints.  
This will allow us to discuss issues that arise when sector decomposition 
is used in realistic calculations.  Sector decomposition 
proceeds in a series of simple steps.
\begin{enumerate}
 \item Split the integration region into two sub-regions;
in the first one, $x>y$, and in the second, $y>x$. The integral 
$I$ is written accordingly as the sum of two terms,
$I = I_1+I_2$.    We have 
   \begin{equation}
    I_1 = \int_{0}^{1} {\rm d}x  \int_{0}^{x} {\rm d}y \frac{x^{\ep}y^{\ep}}{(x+y)^2} F_J[s_{ab}(x,y)],~~~~
    I_2 = \int_{0}^{1} {\rm d}y  \int_{0}^{y} {\rm d}x \frac{x^{\ep}y^{\ep}}{(x+y)^2} F_J[s_{ab}(x,y)].
   \end{equation}
 \item Remap each integration to the unit hypercube.  
In $I_1$, make the change $y^{'}=y/x$; in $I_2$, 
set $x^{'}=x/y$.  Performing these changes of variables
(and rewriting $x^{'} \rightarrow x,y^{'} \rightarrow y$ 
for notational ease), we obtain
   \begin{equation}
    I_1 =\int_{0}^{1} {\rm d}x{\rm d}y \frac{x^{-1+2\ep}y^{\ep}}{(1+y)^2} F_{J}[s_{ab}(x,xy)],~~~~
    I_2 =\int_{0}^{1} 
{\rm d}x{\rm d}y \frac{y^{-1+2\ep}x^{\ep}}{(1+x)^2} F_{J}[s_{ab}(xy,y)].
\label{eq43}
   \end{equation}
 \item The singularities in $I_1$ and $I_2$ are now in a factorized form, and can be extracted with an 
   expansion in plus distributions.
 \item If a singularity appears in the $x \rightarrow 1$ limit, as in 
   \begin{equation}
    I = \int_{0}^{1} {\rm d}x{\rm d}y \frac{(1-x)^{\ep}y^{\ep}}{(1-x+y)^2},
   \end{equation}
make the variable change $x^{'}=1-x$ to map the 
singularity to $x^{'}=0$, and then apply the same  steps as above.
\end{enumerate}

There are several features of this process that should be noted.  It is very simple to program a computer to perform 
this routine.  There are three operations that are performed: 
first, a search for locations of possible singularities of the 
integrand; second, a variable substitution of the form
$y \rightarrow xy$, as in $I_1$; third,  
a factorization such as $(x+xy)^2 = x^2(1+y)^2$ in $I_1$ in order to find 
the overall power of $x$.  These operations 
 can be simply performed using symbolic manipulation programs such as 
MAPLE or MATHEMATICA, as can the substitution of 
Eq.(\ref{eq:plusexpansion}) needed for the plus distribution expansion.  
Another feature to notice is that the mappings between 
the variables $\lambda_1$ and the invariant masses $s_{ab}$ 
are different in each sector, as denoted by the different measurement functions $F_{J}[s_{ab}(x,xy)]$ and $F_{J}[s_{ab}(xy,y)]$ in Eq.(\ref{eq43}).  
If we build an event generator according to the probability distribution in $\lambda_i$, we must account for this different kinematics 
in each sector.  Finally, applying this technique increases
the expression size each time a decomposition is performed, 
so it is best to limit the number of sectors by choosing 
a phase-space parameterization that factorizes as 
many singularities as possible.
   
There are several other features that we include in our computer routine for sector decomposition.
\begin{itemize}
 \item It is useful to rotate the external momenta in order to change terms in the matrix element to a factorized form.  
  For example, in the partonic channel $g(p_1) + g(p_2) \to H(p_h) + g(p_3) + g(p_4)$, we can perform the rotations 
  $p_1 \leftrightarrow p_2$ and $p_3 \leftrightarrow p_4$.  Consider a term in the matrix element of the form 
  $F_{J}(p_1,p_2,p_3,p_4,p_h)/s_{13}/s_{1h}$, where we have described the measurement function arguments using the external momenta.  
  In the rapidity parameterization, this term has a line singularity arising from $1/s_{13}$, as discussed below 
  Eq.(\ref{rapinvm}).  However, under the 
  rotation $p_1 \leftrightarrow p_2$, it becomes $F_{J}(p_2,p_1,p_3,p_4,p_h)/s_{23}/s_{2h}$, which is in a factorized 
  form.  As long as we account for the rotation in $F_J$, which describes the kinematics, this is permissible.
 \item When an integral requires sector decomposition, and is singular in both limits $\lambda \rightarrow 0,1$, we 
  separate these limits by splitting the integration into the two regions $[0,1/2]$ and $[1/2,1]$.  We then remap each 
  region to the range $[0,1]$.  We find that this decreases the analytical complexity of the result, and 
  improves the numerical precision.  As an example, consider the integral
  \begin{equation}
   I = \int_{0}^{1} {\rm d}x{\rm d}y \frac{x^{\ep}y^{\ep}(1-x)^{-1+\ep}}{(x+y)^2}.
  \end{equation}
  We will evaluate this integral in two ways: (1) by splitting the $x$-integration as described above, and (2) by 
  directly applying the algorithm of sector decomposition. For simplicity, we suppress the measurement function in this example.
  Using the first method, we split the $x$-integration and derive $I = I^{(0)}+I^{(1)}$, with
  \begin{equation}
   I^{(0)} = \int_{0}^{1} {\rm d}x{\rm d}y \, 2^{1-\ep} \frac{x^{\ep}y^{\ep}(1-\frac{x}{2})^{-1+\ep}}{(x+2y)^2},~~~~
   I^{(1)} = \int_{0}^{1} {\rm d}x{\rm d}y \, 2^{1-\ep} \frac{y^{\ep}x^{-1+\ep}(1-\frac{x}{2})^{\ep}}{(2-x+2y)^2},
  \end{equation}
where we have changed $x \to 1-x$ in $I^{(1)}$.
The first integral requires a sector decomposition in the limit 
$x \rightarrow 0,y\rightarrow 0$, while the second one is already 
in a factorized form; we obtain 
  $I = I^{(0)}_{x}+I^{(0)}_{y}+I^{(1)}$, with 
  \begin{equation}
   I^{(0)}_{x} = \int_{0}^{1} {\rm d}x{\rm d}y \, 2^{1-\ep} \frac{x^{\ep}y^{-1+2\ep}(1-\frac{xy}{2})^{-1+\ep}}{(2+x)^2},~~~~ 
   I^{(0)}_{y} = \int_{0}^{1} {\rm d}x{\rm d}y \, 2^{1-\ep} \frac{y^{\ep}x^{-1+2\ep}(1-\frac{x}{2})^{-1+\ep}}{(1+2y)^2}.
  \end{equation}
  All singularities have been separated, and the expressions can be expanded in $\ep$.  The important point to notice is 
  that except for the terms that must be expanded in distributions ($y^{-1+2\ep}$ in $I^{(0)}_{x}$, $x^{-1+2\ep}$ in $I^{(0)}_{y}$, 
  and $x^{-1+\ep}$ in $I^{(1)}$), all other terms in the integrands are finite throughout the entire $(x,y)$ plane.  If we instead 
  directly apply the sector decomposition algorithm to $I$ without first splitting the $x$-integration region, we obtain 
  $I = I_x+I_y$, with
  \begin{equation}
   I_x = \int_{0}^{1} {\rm d}x{\rm d}y \frac{x^{\ep}y^{-1+2\ep}(1-xy)^{-1+\ep}}{(1+x)^2},~~~~
   I_y = \int_{0}^{1} {\rm d}x{\rm d}y \frac{y^{\ep}x^{-1+2\ep}(1-x)^{-1+\ep}}{(1+y)^2}.
  \end{equation}
  In addition to the components that must be expanded in distributions, the term $(1-xy)^{-1+\ep}$ in $I_x$ is singular in the 
  limit $x,y \rightarrow 1$.  Although this type of singularity is integrable, we find that it can lead to numerical 
  instabilities when combined with
parton distribution functions
 and phase-space constraints, particularly when this region of the 
  integration contributes strongly to the result.  It is best to avoid 
  these terms using the split described above.  We note that when we use this split, we can choose to map $x \rightarrow 1$ 
  singularities to $x^{'} \rightarrow 0$ using the variable change $x^{'} = 2(1-x)$ in the $[1/2,1]$ region.
\end{itemize}

Now that we have discussed in detail all the elements that enter our sector decomposition routine, we can present 
our algorithm for factorizing entangled singularities for a given term.
\begin{enumerate}
 \item First, check to see if the term can be rotated into a factorized form.  This can be done by establishing 
  a priority for each denominator structure.  In the example given above, $1/s_{23}/s_{2h}$ would be given a 
  high priority, while $1/s_{13}/s_{1h}$ would be given a low priority.  Perform all possible rotations on a given
  term, and check to see if any of them give a high priority integral.  If so, then we have a factorized form, 
  and we can expand in distributions as in Eq.(\ref{eq:plusexpansion}), and go to the next term.
 \item Split each $\lambda_i$ integration into the two ranges $[0,1/2]$ and $[1/2,1]$, in order to separate 
  $\lambda_i \rightarrow 0,1$ singularities.  This can either be done for all $\lambda_i$ automatically, or for 
  a given denominator structure we can only split integration regions for those $\lambda_i$ that have 
  both $\lambda_i \rightarrow 0$ and $\lambda_i \rightarrow 1$ singular limits.  We will assume in the remainder 
  of the discussion that the first has been done, so that all singularities occur as $\lambda_i \rightarrow 0$.
 \item Pick two $\lambda$ variables, {\it e.g.}, $\lambda_1$ and $\lambda_2$, and check whether the term 
  needs sector decomposition in these two variables.  A term needs sector decomposition if the following 
  conditions are met: (1) it doesn't vanish as $\lambda_1 \rightarrow 0$; (2) it doesn't vanish as $\lambda_2 \rightarrow 0$;
  (3) it vanishes as both $\lambda_1,\lambda_2 \rightarrow 0$.  This is a simple check to program in symbolic 
  manipulation packages.  It follows from the expressions for the invariant 
masses given in Section II that we can restrict ourselves 
to singular regions where two of the $\lambda_i$ variables vanish; there is no need to consider 
triple or quadruple singular limits.

 \item If a term doesn't need sector decomposition, return to step 3 and pick another pair of $\lambda_i$.  If it does, then 
  apply the technique discussed above.  This gives two sectors.  Pick the first sector, and begin at step 3 for this term.

 \item Eventually, a term will have no pair of variables 
that requires sector decomposition.
This term is then in a factorized form, 
  and we can extract phase-space singularities by expanding in 
plus distributions.

 \item Repeat steps 1-5 for all terms.
\end{enumerate}

We must now discuss how to handle line singularities.  These arise when a singularity is mapped to an edge of phase space 
rather than just a point because of a specific choice of parameterization.  Our basic method will be to reduce them to an 
entangled form, and then apply the method detailed above.  These singularities appear in the more complicated topologies, 
with a larger number of invariant masses in the denominator.  In 
simpler topologies, such as the $1/s_{13}/s_{1h}$ example discussed above, they can be rotated away.  
It is difficult to automate this reduction as thoroughly as 
the factorization of entangled singularities.  
We will therefore discuss in detail the only case needed for Higgs hadroproduction, which is the 
$1/s_{13}$ line singularity in the rapidity parameterization.  We can avoid the $1/s_{34}$ energy 
parameterization singularity, as we will show in the next section.  We will discuss how to handle the 
topology $1/s_{13}/s_{24}/s_{34}$, which is the simplest topology that can not be rotated away from a line 
singularity form.  For notational simplicity, we will suppress any possible numerator and the measurement function $F_J$ for this structure.

We begin by combining the expressions for $s_{13}$, $s_{24}$, and $s_{34}$ in Eq.(\ref{rapinvm}) with the phase 
space in Eq.(\ref{rapparam}).  We obtain
\begin{eqnarray}
I &=& \frac{N}{(1-z)^4} \int_{0}^{1} {\rm d}\lambda_i \, \left[(1-\lambda_1)(1-\lambda_1 K_m/K_p)\right]^{-\epsilon}
 \left[\lambda_1(1-\lambda_2)\right]^{-1-\ep} \lambda_2^{-\ep} \left[\lambda_3(1-\lambda_3)\right]^{-1-2\ep}
 \nonumber \\ & & \times \left[\lambda_4(1-\lambda_4)\right]^{-\epsilon-1/2}  K_p r 
 \left[K_p r/(1+u)^2\right]^{\epsilon} \left[1-\frac{\lambda_1 K_m}{r(1+z)}\right] \frac{1}{\hat{s}_{13}(\lambda_4)},
\end{eqnarray}
with
\begin{eqnarray}
\hat{s}_{13}(\lambda_4) &=& \lambda_4 (\hat{s}^{+}_{13}-\hat{s}^{-}_{13}) +\hat{s}^{-}_{13}, \nonumber \\ 
\hat{s}^{\pm}_{13} &=& A_1+A_2+ \pm 2\sqrt{A_1A_2}
\end{eqnarray}
and $A_1,A_2$ defined in Eq.(\ref{Adef}).  We note that the line singularity occurs when $\lambda_4=0$ and $A_1=A_2$.  The 
first step is to separate these two requirements.  To do so, we make use of the freedom to map singularities 
nonlinearly to the unit hypercube; so far, we have only used linear transformations such as 
the one used for $s_{13}$: $s_{13}=-\lambda_4(s^{+}_{13}-s^{-}_{13})-s^{-}_{13}$.  The nonlinear mapping 
is achieved by making the additional variable change
\begin{equation}
 \hat{\lambda}_{4} = \frac{\hat{s}^{-}_{13}\times (1-\lambda_4)}{\hat{s}_{13}(\lambda_4)} \rightarrow 
 \lambda_{4} = \frac{\hat{s}^{-}_{13}\times (1-\hat{\lambda}_4)}{\hat{s}_{13}(\hat{\lambda}_4)}.
\end{equation}
The integral becomes
\begin{eqnarray}
I &=& \frac{N}{(1-z)^4} \int_{0}^{1} {\rm d}\lambda_i \, \left[(1-\lambda_1)(1-\lambda_1 K_m/K_p)\right]^{-\epsilon}
 \left[\lambda_1(1-\lambda_2)\right]^{-1-\ep} \lambda_2^{-\ep} \left[\lambda_3(1-\lambda_3)\right]^{-1-2\ep}
 \nonumber \\ & & \times \left[\lambda_4(1-\lambda_4)\right]^{-\epsilon-1/2}  K_p r 
 \left[K_p r/(1+u)^2\right]^{\epsilon} \left[1-\frac{\lambda_1 K_m}{r(1+z)}\right] \left[\hat{s}_{13}(\lambda_4)\right]^{2\ep}
 \nonumber \\ & & \times |\lambda_1(1-\lambda_2)(1+u)^2-(1-\lambda_1)\lambda_2 r(K_p-\lambda_1 K_m)|^{-1-2\ep},
\end{eqnarray}
where we have relabeled $\hat{\lambda}_4 \rightarrow \lambda_4$ for notational simplicity.  The $A_1=A_2$ line has 
been made manifest in the absolute value in the last line of this equation.  As noted in Eq.(\ref{Aphys}), we can 
describe this line by the parameter $\lambda^{s}_2$.  If we split the $\lambda_2$ integration into the two 
regions $[0,\lambda^{s}_{2}]$ and $[\lambda^{s}_{2},1]$, we can force this singularity to always occur at a point 
on the boundary of phase space; this will reduce it to 
an entangled form, amenable to sector decomposition.  We 
denote the integrals over these 
two regions by $I_0$ and $I_1$, so that $I=I_0 +I_1$.  In $I_0$ we perform the variable change 
$\lambda^{'}_{2}=\lambda_{2}/\lambda^{s}_{2}$, while in $I_1$ 
we use $\lambda^{'}_{2}=(\lambda_2-\lambda^{s}_{2})/(1-\lambda^{s}_{2})$.  
We obtain (after relabeling $\lambda^{'}_{2} \rightarrow \lambda_2$)
\begin{eqnarray}
I_0 &=& \frac{N}{(1-z)^4} \int_{0}^{1} {\rm d}\lambda_i \, \left[(1-\lambda_1)(1-\lambda_1 K_m/K_p)\right]^{-\epsilon}
  \left[\lambda_1(1-\lambda_2)\right]^{-1-2\ep} \lambda_2^{-\ep} \left[\lambda_3(1-\lambda_3)\right]^{-1-2\ep}\nonumber \\ 
& & \times \left[\lambda_4(1-\lambda_4)\right]^{-\epsilon-1/2}  K_p r 
  \left[K_p r/(1+u)^4\right]^{\epsilon} \left[1-\frac{\lambda_1 K_m}{r(1+z)}\right] \left[A^{(0)}_{1}\right]^{-1-\ep}
  \nonumber \\ & & \times \left[A^{(0)}_{1}+A^{(0)}_{2}+2(2\lambda_4-1)\sqrt{A^{(0)}_{1}A^{(0)}_{2}}\right]^{2\ep}, \nonumber \\
I_1 &=& \frac{N}{(1-z)^4} \int_{0}^{1} {\rm d}\lambda_i \, \left[(1-\lambda_1)(1-\lambda_1 K_m/K_p)\right]^{-1-2\epsilon}
  \left[\lambda_1(1-\lambda_2)\right]^{-1-\ep} \lambda_2^{-1-2\ep} \left[\lambda_3(1-\lambda_3)\right]^{-1-2\ep}
  \nonumber \\ & & \times \left[\lambda_4(1-\lambda_4)\right]^{-\epsilon-1/2} \left[K_pr(1+u)^2\right]^{-\ep} 
  \left[1-\frac{\lambda_1 K_m}{r(1+z)}\right] \left[A^{(1)}_{1}\right]^{-\ep}
  \nonumber \\ & & \times \left[A^{(1)}_{1}+A^{(1)}_{2}+2(2\lambda_4-1)\sqrt{A^{(1)}_{1}A^{(1)}_{2}}\right]^{2\ep}, \nonumber \\
A^{(0)}_{1} &=& \lambda_1(1-\lambda_2)(1+u)^2+(1-\lambda_1)r(K_p-\lambda_1 K_m), \nonumber \\ 
A^{(0)}_{2} &=& (1-\lambda_1)\lambda_2 r(K_p-\lambda_1 K_m), \nonumber \\ 
A^{(1)}_{1} &=& \lambda_1(1+u)^2+(1-\lambda_1)\lambda_2 r(K_p-\lambda_1 K_m), \nonumber \\ 
A^{(1)}_{2} &=& \lambda_1(1-\lambda_2)(1+u)^2.
\end{eqnarray}
The terms with $A^{(i)}_{1}$ and $A^{(i)}_{2}$ contain singularities entangled in $\lambda_1$ and $\lambda_2$ that can 
be extracted with sector decomposition.  Although the above expressions are somewhat complicated, the simple, automated 
algorithm described above can separate and extract all singularities with minimal human intervention.  
We note that after splitting $I_0$ and $I_1$ into the ranges $[0,1/2]$ and 
$[1/2,1]$ for $\lambda_1$, $\lambda_2$, $\lambda_3$, and factorizing all singularities, we obtain twenty sectors for this 
topology, each with a different mapping between the $s_{ab}$ and the $\lambda_i$.  It is clearly advantageous to 
avoid line singularities by rotating them into entangled or factorized forms when possible.

\section{Topologies for Higgs hadro-production}
\label{sect:top}

We have discussed how to extract singularities from all terms that appear in the double real emission 
corrections to Higgs production at NNLO.  We must now explain what singular structures appear in the 
matrix elements, what parameterizations we use to deal with them, and how the numerator structures are handled.  
We consider the topologies that appear in $g(p_1) + g(p_2) \to H(p_h) + g(p_3) + g(p_4)$; all other partonic 
channels contain only a subset of these topologies.

There is a certain amount of ambiguity in how to map terms in the matrix elements to different topologies.  For example, 
the matrix elements can contain two sets of terms with denominators $1/s_{2h}/s_{34}$ and $1/s_{1h}/s_{23}/s_{34}$.  There 
are two possible ways of dealing with these terms.  We can treat them separately, or we can rotate $p_1 \leftrightarrow p_2$ 
in the first and combine them over a common denominator.  Each approach offers potential advantages.  Combining them 
allows us to extract singularities from only a single structure, $1/s_{1h}/s_{23}/s_{34}$, which reduces both the 
number of singular regions in $\lambda_i$ space, and the expression size.  The first term is singular as $\lambda_3 \rightarrow 1$, while 
the second is singular as $\lambda_3 \rightarrow 0$; the combined expression is singular only when $\lambda_3 \rightarrow 0$.  
However, combining these terms can complicate numerical cancellations between terms in the matrix elements.  As 
noted, to combine these terms requires rotating $p_1 \leftrightarrow p_2$ in $1/s_{2h}/s_{34}$, which 
changes the $\lambda_3 \rightarrow 1$ limit to $\lambda_3 \rightarrow 0$.  Suppose there is a numerical 
cancellation between the $1/s_{2h}/s_{34}$ term and another component that occurs locally in $\lambda_i$-space 
when $\lambda_3 \rightarrow 1$.  After rotation, this cancellation will only occur globally after integrating 
over $\lambda_3$.  

The point of this example is to show the types of possibilities 
that exist when we map the matrix elements to topologies.  It is not clear to us at the moment
 which way is ``better'': 
which leads to smaller expressions and to more 
numerically stable results, {\it etc.}  
In the discussion that follows, we focus on the parameterization-
independent features of the possible mappings.
\begin{itemize}
 \item We will note which  topologies are necessarily of an entangled or a line-singular form.
 \item We will indicate which topologies are ``soft'', {\it i.e.}
singular in the  $z \rightarrow 1$ limit.  
  These only occur for the partonic channels with a $gg$ initial state: $gg \rightarrow ggH$ and 
  $gg \rightarrow q\bar{q}H$.
 \item We will discuss the topologies with ``quadratic'' singularities; we will explain later in detail what this means.
\end{itemize}
We will also try to explain what techniques we found useful in organizing our calculation, and will attempt to 
discuss the possible highlights and disadvantages of different methods.

We first discuss how we deal with the numerator structures of each 
topology.  Each term that appears 
in the matrix elements has the form 
\begin{equation}
\frac{{\cal N}[s_{ab}(\lambda_i)]}{s_{ab}s_{cd}\ldots},
\end{equation}
where the numerator ${\cal N}$ contains 
polynomials of the invariant masses in addition to measurement functions $F_J$ with 
various arguments.  The denominator characterizes the topology because 
it is insensitive to the exact particle content of the theory, and 
only depends on the global structure of the process under consideration.
Identical topologies will appear for all $2 \to 3$ processes with 
one massive particle in the final state. On the contrary, numerators 
are process specific; for example, in a theory with only scalar 
particles, all the numerators are equal to one.  It is therefore advantageous to 
treat the numerators and denominators of the matrix elements separately.

One way to deal with the numerators in a numerical program 
is to explicitly substitute the expressions 
for the $s_{ab}$ in terms of the $\lambda_i$ for each topology, so that everything in our Fortran code is 
written in terms of $\lambda_i$.  However, we found that it is better 
to define auxiliary functions $s_{ab}(\lambda_i)$ 
to 
keep ${\cal N}$ written in terms of the $s_{ab}$ in our numerical 
routines.  We 
introduce the explicit formulae for the $s_{ab}$ only for 
the denominators, which is required in order to 
extract singularities.  To show what appears in our 
numerical code, consider the topology 
${\cal N}[s_{ab}(\lambda_2,\ldots)]/s_{23}$.  This is singular as $\lambda_2 \rightarrow 0$, as is clear from 
Eq.(\ref{rapinvm}), so we get terms such as ${\cal N}[s_{ab}(\lambda_2,\ldots)]/[\lambda_2]_+$.  What we write to our 
code is 
\begin{equation}
\frac{{\cal N}[s_{ab}(\lambda_2,\ldots)]-{\cal N}[s_{ab}(0,\ldots)]}{\lambda_2}.
\end{equation}
The terms $s_{ab}(\lambda_2,\ldots)$ and $s_{ab}(0,\ldots)$ are obtained numerically by calls to a Fortran 
function.  This 
drastically reduces expression sizes; for example, referring to Eq.(\ref{rapinvm}), we see that if a term 
has $s_{13}$ in the numerator, it is much cleaner to keep it written as $s_{13}(\lambda_i)$ rather than explicitly 
introduce the lengthy functional form.  Since all topologies 
depend upon a small set of $s_{ab}$, optimization of the 
code is appreciable.  As mentioned above, the denominator structures are universal; the 
same ones appear in Higgs hadroproduction, electroweak gauge boson production, 
$b\bar{b} \rightarrow H$ production in supersymmetric theories, and a 
host of other phenomenologically 
interesting $2 \rightarrow 1$ processes.  Structuring our numerical code 
by treating the denominators and numerators of the matrix elements 
differently allows us to study other processes simply.

We are now ready to discuss the topologies that appear in Higgs hadroproduction.  We first introduce some notation.  
We refer to a topology with four invariant masses as a ``highest-level'' topology, as this is the maximum number 
of scalar products that can occur in the denominator for the double real emission contributions.  When we remove 
an invariant mass from a topology, we refer to this as a ``sub-topology'' of the original term.  We remind the 
reader of the three classes of diagrams we introduced in the section on phase-space parameterizations.  We will 
refer to them as $C_1,C_2,C_3$, and interferences between them as $C_1 \times C_2$, {\it etc.}  Only topologies 
independent under rotation will be discussed; for example, if we present $1/s_{13}/s_{24}$, we will not 
consider $1/s_{14}/s_{23}$.
\begin{itemize}
 \item The only highest-level topology that necessarily contains a line singularity is $1/s_{13}/s_{24}/s_{34}/s_{1h}$.  
   It occurs in $C_2 \times C_3$.  It is also a soft topology, and therefore begins at $1/\ep^4$.  
   We found it best to map this term into the rapidity 
   parameterization; with this choice, we obtain 20 sectors, as we noted in our discussion in the previous section on 
   line singularities.  If we had chosen instead the energy parameterization, more sectors would be required to 
   factorize the singularities present in $s_{1h}$.  This is the most difficult topology to evaluate numerically; 
   the singular $z \rightarrow 1$ limit creates a large number of plus distributions at the finite level.  It has 
   a subtopology $1/s_{13}/s_{24}/s_{34}$ that also has a line singularity; its remaining subtopologies do not.
 \item There are several highest-level topologies that necessarily contain entangled singularities.  The two most 
   complicated of these are $1/s_{13}/s_{23}/s_{1h}/s_{2h}$ and $1/s_{13}/s_{24}/s_{1h}/s_{2h}$, which appear in 
   $C_1 \times C_3$.  They are both soft, and begin at $1/\ep^4$.  If we evaluate them using the energy 
   parameterization, we can avoid line singularities associated with $s_{13}$.
 \item The other highest-level topology that is necessarily of entangled form is $1/s_{13}/s_{23}/s_{24}/s_{1h}$ 
   (another one that requires a special 
   discussion will be considered later).  It is not soft, and so only begins at $1/\ep^3$.  It
   appears in $C_1 \times C_3$.  We can again avoid line singularities by using the energy parameterization.
\end{itemize}

We must now discuss topologies with ``quadratic'' singularities.  To explain this concept, we will discuss the 
simplest example where quadratic singularities occur.  Consider the sub-topology $1/s_{23}/s_{34}/s_{1h}$ in the rapidity parameterization.  
When we combine the expressions in Eq.(\ref{rapinvm}) with the phase space in Eq.(\ref{rapparam}), we find that 
this topology scales as $\lambda_{3}^{-2-2\ep}$; the topology is quadratic in $\lambda_3$.  Singularities this 
strong are unphysical.  If we were to regulate them with a mass $m$ rather than with $\ep$, we would find 
a $1/m^2$ rather than a ${\rm ln}(m^2)$ singularity, which can not occur.  Furthermore, the plus distribution 
expansion in Eq.(\ref{eq:plusexpansion}) is only valid for linear singularities.  When we find quadratic singularities, 
we always have a term in the numerator of the topology that renders the behavior acceptable.  In this 
example, we must find that the numerator scales as $\lambda_3$.  However, finding this scaling requires 
introducing the explicit expressions for the invariant masses in terms of the $\lambda_i$.  We want to do as 
little of this as possible in order to keep term sizes small.   Topologies with quadratic singularities 
in $\lambda_3$ are the simplest to handle.  We map all such cases to the rapidity parameterization.  Referring 
to Eq.(\ref{rapinvm}), we observe that all the invariant masses have a simple, overall scaling in $\lambda_3$, 
together with a more complicated term.  For example, $s_{1h} \sim \lambda_3$, $s_{1h} \sim (1-\lambda_3)$ 
$s_{13} \sim (1-\lambda_3)$, {\it etc.}  We find that introducing this overall factor is sufficient to 
cancel quadratic singularities; the terms in brackets that appear in these scalar products never enter the 
cancellation, and can be kept implicit.  This allows us to keep the term sizes small.

The remaining quadratic topologies, {\it i.e.}, those with a quadratic singularity in a variable other then $\lambda_3$, 
require a slightly more involved procedure.  There is typically a 
combination of invariant masses in the numerator for which we must introduce an explicit parameterization in order 
to regulate the singularity.  
\begin{itemize}
 \item In the interference $C_1 \times C_1$, we find the topology $1/s^{2}_{13}/s^{2}_{24}$.  This is 
  factorizable in the energy parameterization; however, referring to Eq.(\ref{eninvm1}), we observe that 
  it scales as $(1-\lambda_2)^{-2-\ep}\lambda_{3}^{-2-\ep}$.  The relevant numerator structure that regulates 
  these quadratic singularities is 
  \begin{equation}
    \left(s_{14}s_{23}-s_{34}\right)^2 \propto (1-\lambda_2)\lambda_3(1-z)^4.
  \end{equation}
  The $(1-z)^4$ factor that appears ensures that this topology is not soft.
 \item Two other similar quadratic topologies that appear are $1/s_{24}^2/s_{1h}^2$ and $1/s_{24}^2/s_{13}/s_{1h}$.  
   The first comes from $C_3 \times C_3$.  It is factorizable in the rapidity parameterization, where we find 
   that it scales as $(1-\lambda_{2})^{-2-\ep}$. The second topology comes from $C_1 \times C_3$, and 
   necessarily contains entangled singularities.  We can avoid a line singularity in the energy parameterization, 
   where we find that it also scales as $(1-\lambda_{2})^{-2-\ep}$.  They are both regulated by the same numerator 
   structure as above, $\left(s_{14}s_{23}-s_{34}\right)^2$.  Neither topology is soft.
 \item The final two quadratic topologies appear in $C_2 \times C_2$: $1/s_{34}^2/s_{1h}^2$ and $1/s_{34}^2/s_{1h}/s_{2h}$.  
   They are factorizable in the rapidity parameterization, where we find that they scale as $\lambda_{1}^{-2-\ep}$.  
   The required numerator structure is 
   \begin{equation}
     (z s_{23}-s_{13}s_{1h}-s_{23}s_{1h}-s_{23})^2 \propto \lambda_1 \lambda_{3}^{2} (1-\lambda_3)^2 (1-z)^4,
   \end{equation}
   which also ensures that the topology is not soft and that 
quadratic singularities in $\lambda_1$  cancel.
\end{itemize}

All of the remaining topologies that are needed for Higgs 
hadroproduction can be mapped directly to 
a factorized form in either the energy or rapidity parameterization.  
We find the following highest-level topologies:
\begin{itemize}
 \item $1/s_{23}/s_{24}/s_{1h}^2$ in $C_3 \times C_3$, which we 
map to the rapidity parameterization;
 \item $1/s_{23}/s_{34}/s_{1h}^2$ in $C_2 \times C_3$, which we 
map to the rapidity parameterization;
 \item $1/s_{23}/s_{34}/s_{1h}/s_{2h}$ in $C_2 \times C_3$, which 
we map to the rapidity parameterization;
 \item $1/s_{13}/s_{14}/s_{23}/s_{24}$ in $C_1 \times C_1$, which 
we map to the energy parameterization.
\end{itemize}
Only the final two topologies in this list are soft.  
We can directly apply the expansion in Eq.(\ref{eq:plusexpansion}) to these 
terms to extract the phase-space singularities.

\section{Results} 

In this Section we describe phenomenological results obtained using 
our calculation.  Our primary goal is to illustrate 
the range  of observables that can be studied using the approach 
discussed in this paper, not to perform a comprehensive 
study of  the Higgs boson signal at the LHC.  
A state-of-the-art phenomenological analysis of the Higgs boson 
signal is required for two reasons. First, it can be used to optimize cuts, 
which is important for enhancing the signal-to-background 
ratio. Second, it is useful for exposing uncertainties in the 
theoretical prediction for the Higgs boson signal, including 
truncation of the perturbative 
expansion at NNLO, choices of the factorization and the renormalization 
scales, and imprecise knowledge of parton distributions and $\alpha_s(M_z)$. Although such a study 
is beyond the scope of this paper, we hope that the examples 
presented below are sufficiently convincing to demonstrate 
 that our approach permits a computation of
{\it arbitrary} observables related to the reaction $pp \to H +X \to 
{\rm Higgs \, decay \, products} +X$. In particular, we present below the 
fully realistic NNLO cross sections
for the Higgs 
boson signal in the di-photon decay channel, where the two photons in the 
final state satisfy all the selection criteria (cuts on photon 
pseudorapidities, transverse momenta, and geometric isolation 
from significant hadronic activity) used by the ATLAS and CMS collaborations.

We remind the reader that we work in the  limit of an infinitely heavy top quark, 
in which the Higgs boson 
coupling to gluons is point-like. However, all the results presented 
in this Section are rescaled by a factor
\be
F(m_t) = \frac{\sigma_{\rm LO}(m_t)}{\sigma_{\rm LO}(m_t = \infty)}.
\label{eq85}
\ee
This rescaling accounts for the effects of the finite top mass exactly for 
LO cross sections, and provides an approximate description of the 
top mass dependence at higher orders. For NLO calculations, this 
is known to be an excellent approximation,
and we expect it to work well at NNLO also.

\begin{figure}[ht]
\centerline{
\epsfig{figure=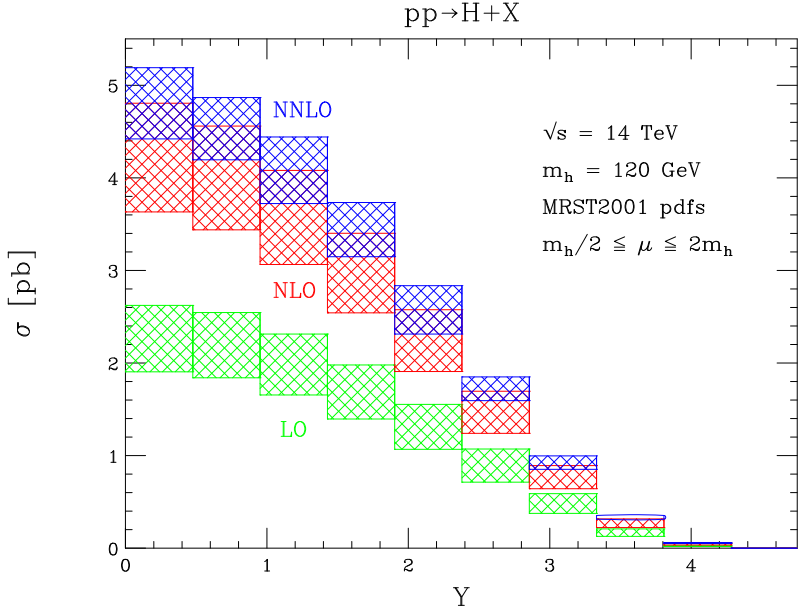,width=3.2in,angle=90}
}
   \caption{\label{fig:hrap}
Bin-integrated Higgs boson  rapidity distribution at the LHC.  The bands indicate 
the scale choice $m_h/2 \leq \mu \leq 2m_h$.
}
\end{figure}

All the results that we present in this Section 
can be divided  into two categories, depending on whether or not 
the  decay of the Higgs boson into two photons is considered. We first 
study the simpler case, in which the Higgs boson is treated as a stable 
particle, and investigate the various kinematic  distributions of the 
Higgs. We then turn to the analysis of the 
Higgs signal in the di-photon channel, and present the NNLO  
cross sections and the distributions in the 
pseudorapidity difference $|\eta^{\gamma,1} - \eta^{\gamma,2}|/2$
and the average transverse momentum $(p_\bot^{\gamma,1} 
+ p_\bot^{\gamma,2})/2$ of the two photons.

The rapidity distribution of the Higgs boson with mass 
$m_h = 120~{\rm GeV}$  is shown in Fig.~\ref{fig:hrap} 
at LO, NLO, and NNLO in QCD perturbation theory. 
The bands are obtained by equating the 
factorization and the renormalization scales $\mu_f = \mu_r = \mu$,  
and then varying $\mu$ in the range $m_h/2 \leq \mu \leq 2 m_h$. We use the 
parton distribution functions (pdfs) 
from  the MRST2001 set \cite{mrst}, and employ {\sf mode 1} as the default mode for pdf evolution.  
Higher order QCD 
corrections to the Higgs rapidity distribution exhibit 
behavior similar to the corrections to the inclusive cross section;
the large increase in ${\rm d}\sigma/{\rm d Y}$ from LO to NLO, 
is {\it not} followed by a similar large increase from NLO to NNLO.
The NNLO corrections are uniform over the central rapidity interval $|Y| < 2$, 
and can therefore be obtained to good approximation by re-scaling the NLO
rapidity distribution by a universal,  rapidity-independent factor.

\begin{figure}[ht]
\centerline{
\epsfig{figure=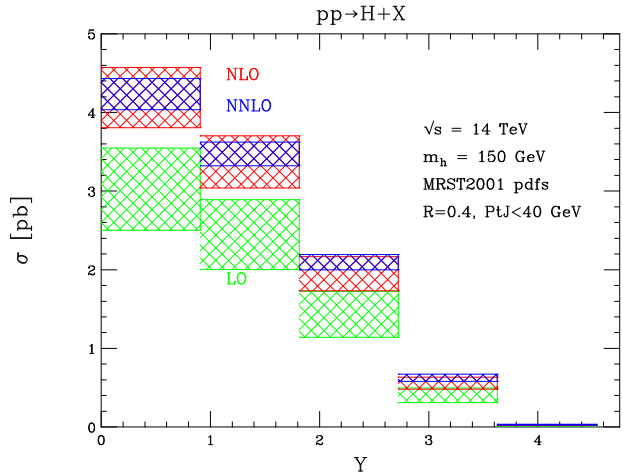,width=3.2in}
}
   \caption{\label{fig:vrap}
Bin-integrated Higgs boson rapidity distributions at the LHC 
with a jet veto applied.
}
\end{figure}

For a Higgs boson heavier than $\sim 140~{\rm GeV}$, the 
decay $H \to W^+W^-$ becomes the dominant decay mechanism, 
making the Higgs branching ratio into two photons quite small; 
consequently, the two photon signal can not
be used as the  primary trigger for the Higgs. Searching 
for the Higgs boson in the $W^+W^-$ decay mode requires 
the introduction of additional cuts to suppress the background  
due to the production and subsequent decay of a pair of top quarks.
Since the hadronic jets in top pair production have, on average, larger 
transverse momenta than hadronic jets in Higgs  
hadroproduction, the significance of the 
Higgs signal can be enhanced by imposing 
a jet veto on the recoiling hadronic system \cite{veto,vetoexp}. 
In Fig.~\ref{fig:vrap} we present 
the rapidity distribution of the Higgs boson with mass 
$m_h = 150~{\rm GeV}$, when all jets in the 
final state of the reaction $pp \to H +X$ are required to have 
transverse momenta smaller than $p^{\rm jet}_{\rm T, veto} = 40~{\rm GeV}$.
The jets are identified with the cone algorithm, using a cone size $R = 0.4$.

It is instructive  to compare both the magnitude of the perturbative 
corrections and the residual scale dependence of the Higgs 
rapidity distributions with and without 
the jet veto. Although Figs.~\ref{fig:hrap},\ref{fig:vrap} 
refer to different masses of the 
Higgs boson, a qualitative comparison is still possible; as 
is known from the NNLO computation of the inclusive cross section, 
the relative magnitude of the NNLO corrections does not depend 
strongly on the Higgs boson mass.
 From Figs.~\ref{fig:hrap},\ref{fig:vrap} 
we observe that the application of the  
jet veto leads to a better perturbative stability of the rapidity 
distribution; the 
perturbative corrections are smaller when the jet veto is applied, and
 there is a complete overlap of the NNLO scale dependence band 
with the NLO band, for most rapidities. This comparison shows that 
large and perturbatively unstable contributions to the Higgs boson 
hadroproduction cross section are related to kinematic configurations 
where the Higgs is produced with large transverse momentum.
This feature does not imply a breakdown of the perturbative 
expansion; it merely reflects the fact that 
for Higgs hadroproduction, the leading order 
partonic process is $gg \to H$, so that at LO 
the Higgs boson is produced with zero transverse 
momentum. Therefore, 
if Higgs boson production with $p_\bot^{\rm higgs} \ne 0$ is considered, 
our NNLO computation includes just two terms in the perturbative expansion 
in the strong coupling constant, and is therefore a NLO computation.

\begin{figure}[ht]
\centerline{
\epsfig{figure=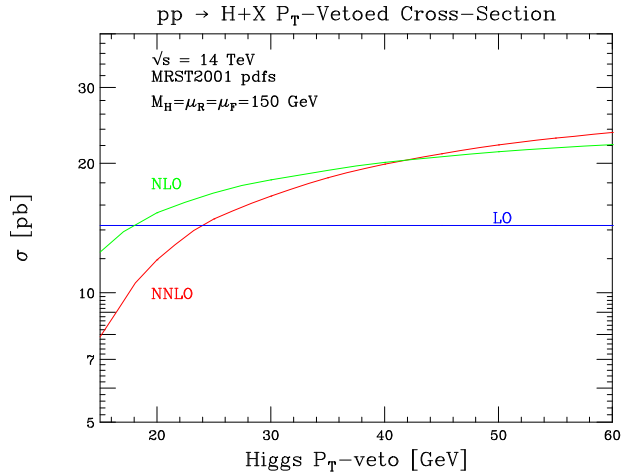,width=3.2in}
}
   \caption{\label{fig:ptnnlo}
The Higgs boson production cross section  at the LHC 
as a function of $p_{\rm T,veto}$.
}
\end{figure}

The excellent convergence observed for the jet-vetoed rapidity distribution 
with $p^{\rm jet}_{\rm T,veto}=40~{\rm GeV}$ is partly accidental. This can 
be seen from  Fig.~\ref{fig:ptnnlo}, where we show the dependence of the 
Higgs boson production cross section on 
the value of a veto on the transverse energy of the Higgs, $p_{\rm T,veto}$. While the LO cross section obviously does 
not depend on $p_{\rm T,veto}$, both the NLO and the NNLO cross sections 
exhibit a significant dependence on this parameter. It is interesting 
to observe that the NLO cross section reaches its asymptotic value much 
faster than the NNLO one; this is related to the fact that the average 
$p_{\rm T}$ of the Higgs boson {\it increases} from NLO to NNLO. For example, 
for $m_h = 150~{\rm GeV}$, the average transverse momenta of the 
Higgs boson at NLO and NNLO are $\langle p_\bot^{\rm NLO} \rangle =37.5~{\rm GeV}$ 
and $\langle p_\bot^{\rm NNLO} \rangle =44.6~{\rm GeV}$. 
It is clear from Fig.~\ref{fig:ptnnlo} 
that the choice  $p_{\rm T,veto} = 40~{\rm GeV}$
minimizes the NNLO QCD corrections 
for $m_h = 150~{\rm GeV}$. However, 
other choices of $p_{\rm T, veto}$ also lead to only small differences 
between the NLO and the NNLO cross sections, indicating an improved convergence  of the perturbative expansion 
for all veto choices.

\begin{figure}[ht]
\centerline{
\epsfig{figure=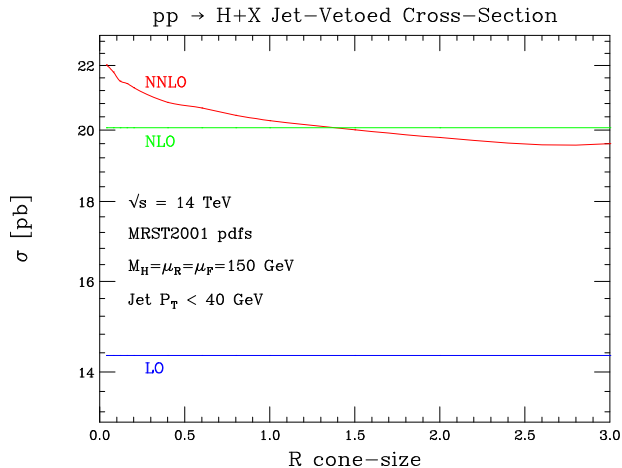,width=3.2in}
}
   \caption{\label{fig:rdep}
The Higgs boson production cross section  at the LHC with a 
jet-veto  of $p_{\rm T,veto}=40~{\rm GeV}$, as a function 
of the cone size $R$. 
}
\end{figure}

Another interesting feature of the NNLO result 
is the dependence of the jet-vetoed 
cross section on the cone size $R$.  At NLO, every jet 
is associated with a massless gluon or quark, so there is no 
cone-size dependence 
of the cross section.  At NNLO, when two partons are close in rapidity and azimuthal angle, 
they are combined into a single jet. Thus, a cone-dependence 
of the observable cross section appears.
As can be seen from Fig.~\ref{fig:rdep}, the cross section decreases when 
the cone size is increased over a large range of $R$; however, 
at large values of $R$ the cross section starts to increase again.
This is a consequence  of the fact that, originally, combining two 
energetic partons into a single jet increases its transverse momentum.
However, after the cone size becomes  so large that two partons 
which are back-to-back in the transverse plane are combined to form a single 
jet, the momentum of such a jet becomes {\it smaller} than the momenta 
of the individual partons. This effect drives an increase in the cross section 
for $R \ge 2.5$.

Since the rapidity of the Higgs boson is not probed, the results 
of Fig.~\ref{fig:rdep} can also be obtained~\cite{veto} by subtracting the NLO contributions 
for $H+1 \mbox{ jet}$ production~\cite{deFlorian:1999zd,Campbell:2000bg} 
from the NNLO total cross section. We have compared our results with 
those in Refs.~\cite{deFlorian:1999zd,Campbell:2000bg}, and have 
found excellent agreement.

We now discuss the Higgs boson signal in $pp \to H+X \to \gamma \gamma +X$.
An interesting observable is the total di-photon 
production cross section, subject to a number of 
cuts on the kinematic variables of the two photons. 
These cuts, designed to enhance the signal over the prompt photon 
background, include restrictions on the photon transverse momenta and 
the rapidities, as well as isolation cuts. 
In particular, the photons are required to have transverse momenta 
$p_\bot^{(1)} \ge 40~{\rm GeV}$ and 
$p_\bot^{(2)} \ge 25~{\rm GeV}$; they must also be produced in the central 
rapidity region $|\eta| <2.5$~\cite{search}. The isolation cuts 
are designed to reduce contamination of the signal by the poorly 
known fragmentation contributions. We require
that a photon candidate does not have
an additional transverse energy  $E_{\rm T}$ which is greater than $ E_{\rm T,min}=15~{\rm GeV}$ 
deposited within a cone around it of radius $R_{\rm is} = 
\sqrt{(\eta-\eta_\gamma)^2 +  (\phi-\phi_\gamma)^2} = 0.4$.  We will refer to this combination 
of cuts as the ``standard photon cuts'' in what follows. 

\begin{table}[htb]
\begin{center}
$$
\begin{array}{||c|c|c||c|c||c|c||}
\hline
&
\multicolumn{2}{|c|}{\sigma_{\rm LO}/Br_{\gamma \gamma},~{\rm pb}}&
\multicolumn{2}{|c|}{\sigma_{\rm NLO}/Br_{\gamma \gamma},~{\rm pb}}&
\multicolumn{2}{|c|}{\sigma_{\rm NNLO}/Br_{\gamma \gamma},~{\rm pb}} \\
\cline{2-7}
m_h,~{\rm GeV}
&\mu=m_h/2 & \mu=2m_h &  \mu=m_h/2 &  \mu=2m_h  
& \mu=m_h/2 & \mu=2m_h 
\\ \hline \hline
110&  16.10&11.67& 31.33 & 
 23.12    &33.53  &27.68  \\ \cline{2-7}
115&   15.59& 11.23  &29.75  &
 21.97
  &31.40  &26.80  \\ \cline{2-7}
120&  15.02& 10.76  & 
28.30  & 
 20.93   & 29.51 & 25.67 \\ \cline{2-7}
125&  14.40&  10.27   &
26.94   & 
 19.92   &28.58  &24.56  \\ \cline{2-7}
130&  13.79   &  9.79 &
 25.52   &
 19.03 
   & 27.92 & 23.29  \\ \cline{2-7}
135&  13.19   &  9.32 &
 24.31    & 
 17.99   &26.04  &22.05 \\ \cline{2-7}
\hline
\end{array}
$$
\vspace*{0.5cm}
\caption{\label{tab:isolation}
The cross section for  $pp \to H + X \to \gamma \gamma +X$, divided 
by the branching ratio $Br(H \to \gamma \gamma)$,
at $\sqrt{s} = 14~{\rm TeV}$ with the standard cuts applied to the photons,
for different values of the Higgs mass.
}
\end{center}
\end{table}

\begin{table}[htb]
\begin{center}
$$
\begin{array}{||c|c|c||}
\hline m_h,~{\rm GeV} & \sigma^{\rm cut}_{\rm NNLO}/\sigma^{\rm inc}_{\rm NNLO} 
 & K^{(2)}_{\rm cut}/K^{(2)}_{\rm inc} \\ \cline{1-3}
110&  0.590 & 0.981 \\ \cline{2-3}
115&  0.597 & 0.968 \\ \cline{2-3}
120&  0.603 & 0.953 \\ \cline{2-3}
125&  0.627 & 0.970 \\ \cline{2-3}
130&  0.656 & 1.00  \\ \cline{2-3}
135&  0.652 & 0.98  \\ \cline{2-3}
\hline
\end{array}
$$
\vspace*{0.5cm}
\caption{\label{tab:kfactors}
Comparisons between the cut and inclusive cross sections for different 
Higgs masses.  The second column contains the ratio of the NNLO cross section 
with the standard cuts over the inclusive cross section, while the third 
column contains the ratio of cut and inclusive results for the $K$-factor $K^{(2)} = \sigma_{\rm NNLO} /  \sigma_{\rm NLO}$.  
We have set $\mu = m_h/2$.
}
\end{center}
\end{table}

In Table~\ref{tab:isolation} we present  the cross section for 
$pp \to H +X \to \gamma \gamma +X$, divided by the 
branching fraction of $H \to \gamma \gamma$, with the 
standard cuts imposed on the photons.  We give results through LO, NLO and NNLO in perturbation 
theory, for different Higgs boson masses and for two choices 
of the renormalization and factorization scales.  The NNLO results are accurate to 
1\%, while the LO and NLO numbers are accurate to 0.1\%.  
The perturbative corrections to the di-photon signal follow the pattern of the corrections to the 
inclusive Higgs production cross section. We note the apparent 
convergence of the di-photon signal, with the NNLO corrections increasing 
the NLO result by up to $\sim 20\%$, depending on the value of $\mu$
and the Higgs boson mass.
The NNLO and NLO scale dependence bands 
overlap significantly, and the NNLO scale dependence is reduced
by a factor of two compared to the NLO dependence.

\begin{figure}[htb]
\centerline{
\epsfig{figure=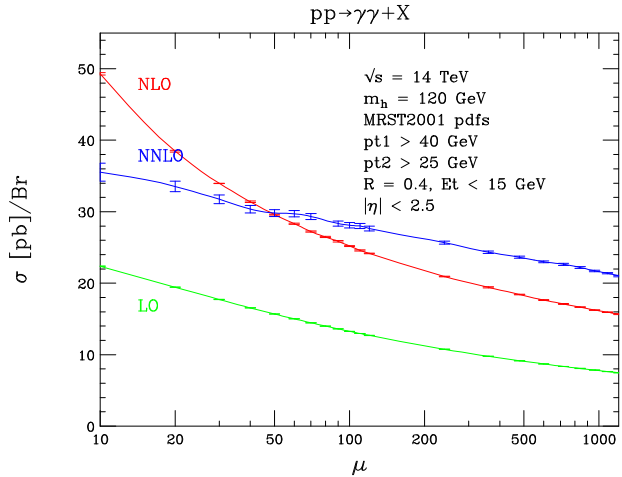,width=3.2in}
}
   \caption{\label{fig:scale_cuts}
The Higgs channel $pp \to H + X \to \gamma \gamma +X$ 
with the standard cuts imposed on the photons, as a function of scale choice.
}
\end{figure}

It is interesting that the NNLO corrections are much smaller when the scale 
$\mu = m_h/2$ is chosen. This is in accord with the observation 
that the typical transverse momentum of the Higgs boson
is parametrically smaller than the Higgs mass $m_h$; 
because of the rapid increase of the gluon pdf at low $x$, the Higgs bosons 
are produced close to threshold. Since the factorization 
scale should be chosen close to a typical transverse momentum of the 
Higgs, selecting $\mu = m_h/2$ expedites the convergence of the 
perturbative expansion.  Fig.~\ref{fig:scale_cuts} 
shows the dependence of the Higgs signal 
on the choice of the scale 
for $m_h = 120~{\rm GeV}$. As expected, we find that the NNLO 
perturbative corrections are small for $\mu \sim 40-50~{\rm GeV}$.  
We note that the threshold resummed results for the 
Higgs hadroproduction cross section \cite{Catani:2003zt} agree very well 
with the fixed order results for smaller scale choices such as $\mu \sim m_h/2$, 
while they differ from the fixed order results by up to several percent for 
larger scale choices.   It appears, from the stability of the 
perturbative series and the agreement with the resummed result, that 
$\mu \sim m_h/2$ is a better scale choice for Higgs hadroproduction.

In Table~\ref{tab:kfactors} we present several comparisons between the inclusive NNLO cross 
section, and the NNLO cross section computed with the standard cuts.  Our goal in 
doing this is two-fold: to understand how the cross section is affected 
by the standard cuts, and to see how well the realistic cut cross section can 
be approximated if only the inclusive NNLO result is known.  We adopt $\mu = m_h/2$ 
for these comparisons.  These results are valid to approximately 2\%.  
In the second column of Table~\ref{tab:kfactors} we have presented the ratio 
of the NNLO cross section with the standard cuts over the inclusive NNLO result.  The 
reduction of the cross section caused by the cuts ranges from 40\% for $m_h < 120$ GeV 
to 30\% for $m_h > 150$ GeV.  We note that most of this reduction comes from 
the $p_{\bot}$ and $\eta$ cuts; the isolation cuts decrease the ratio by 
less than 3\%.  This is expected; there is no cross section enhancement when 
a parton is emitted along the photon direction, so this phase-space region 
contributes minimally to the total result.  We note that the cuts become less 
effective at larger $m_h$, {\it i.e.}, the ratio increases.  For larger 
Higgs masses, the average photon 
$p_{\bot}$ increases, and therefore more events 
pass the cuts. 

In the third column of Table~\ref{tab:kfactors} we present the ratio of the 
$K$-factor $K^{(2)} = \sigma_{\rm NNLO} /  \sigma_{\rm NLO}$.  This is interesting 
for the following reason.  Suppose only the differential
NLO cross section and the 
inclusive NNLO result are known.  The best approximation for the 
exact NNLO differential result would then be 
$d\sigma^{\rm approx}_{\rm NNLO} = d\sigma_{\rm NLO} \times K^{(2)}_{\rm inc}$, where 
$K^{(2)}_{\rm inc}$ is defined with the inclusive cross sections.  Calculating the 
cut cross section with this distribution gives $\sigma^{\rm approx,cut}_{\rm NNLO} = \sigma^{\rm cut}_{\rm NLO} \times 
K^{(2)}_{\rm inc}$.  The ratio of this result with the exact NNLO cross section with the 
standard cuts imposed is 
$\sigma^{\rm cut}_{\rm NNLO}/\sigma^{\rm approx,cut}_{\rm NNLO} = K^{(2)}_{\rm cut} / K^{(2)}_{\rm inc}$; the 
deviation of this ratio from unity measures  the error made by 
using $d\sigma^{\rm approx}_{\rm NNLO}$ to approximate the actual 
differential cross section at NNLO.  We see that this deviation is less 
than about 5\%. 

\begin{figure}[htb]
\centerline{
\epsfig{figure=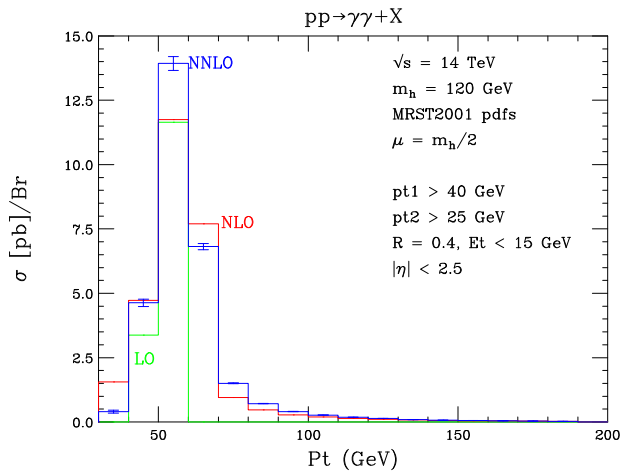,width=3.2in}
}
   \caption{\label{fig:ptsym}
The $p_\bot = (p_\bot^{\gamma,1} + p_\bot^{\gamma,2})/2$ distribution 
for the di-photon Higgs signal at the LHC.}
\end{figure}

 In order to optimize the experimental cuts, it is desirable to 
have a good understanding of the kinematic distributions of the 
photons, since they can provide good discriminators between 
the signal and the background. While we do not 
discuss cut optimization in this paper, we present two differential 
distributions that illustrate the range of observables 
that can be studied using 
our calculation. In Fig.~\ref{fig:ptsym}, the 
$p_\bot = (p_\bot^{\gamma,1} + p_\bot^{\gamma,2})/2$ distribution 
is shown for $m_h = 120~{\rm GeV}$. We observe large perturbative 
corrections close to the kinematic boundary at leading order,
$p_\bot < m_h/2 = 60~{\rm GeV}$, where resummation of large 
logarithms is required.  However, the presence of a large peak near the 
LO kinematic boundary appears to be a reliable result, as it appears without 
drastic modification at both 
NLO and NNLO.  Since the background should not contain any such feature, 
this is potentially a useful discriminating variable.

\begin{figure}[htb]
\centerline{
\epsfig{figure=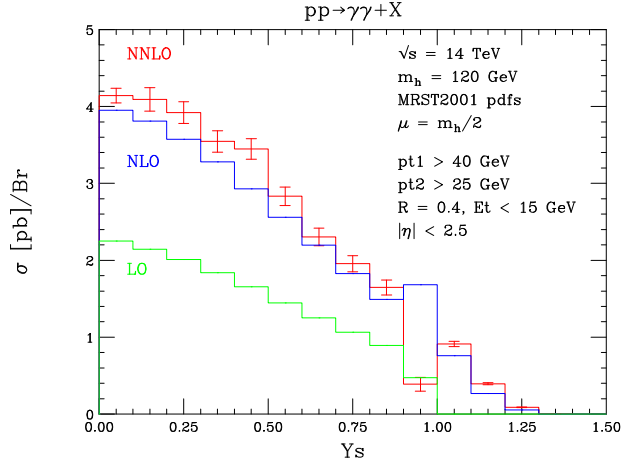,width=3.2in}
}
   \caption{\label{fig:etadiff}
The distribution in the pesudorapidity difference of the 
two photons, $Y_{s}=|\eta_1^{\gamma} - \eta_2^{\gamma}|/2$. }
\end{figure}

In Fig.~\ref{fig:etadiff}, 
we present the distribution of the pseudorapidity difference
$Y_s =|\eta^{\gamma,1} - \eta^{\gamma,2}|/2$ between the two photons.
This distribution is interesting since a similar distribution 
from the prompt photon production background is flatter; this 
information 
can be used to enhance the statistical significance of the Higgs 
signal \cite{lance1}. 
From Fig.~\ref{fig:etadiff} we see that the peak 
at $|\eta^{\gamma,1} - \eta^{\gamma,2}| = 0$ is
also present when the NNLO effects are included. 

The results presented in this Section are for the di-photon 
Higgs signal. There are other Higgs decay modes that are of 
significant interest. In particular, for moderately heavy Higgs bosons, 
decays into $ZZ \to 4l$ and 
$W^+W^- \to l^+l^-\nu_l \bar \nu_l$ might provide suitable channels for  discovery. 
Since our calculation retains all the information about the Higgs boson kinematics, 
it is in principle straightforward to include Higgs decays 
into arbitrary final states. However, in reality, some care must 
be exercised to generate the final-state decay efficiently, 
especially for decays  with high multiplicities and sophisticated 
cuts. We plan on adding additional Higgs decay 
channels to our code in the future.

\section{Description of the {\tt FORTRAN} code}

In this Section we describe a {\tt FORTRAN} program,\hname, which 
we have written to obtain the results described in the previous 
Section. As can be seen from the examples presented there, 
\hname ~computes the cross section for Higgs boson production 
in hadronic collisions through NNLO in perturbative QCD 
in the presence of an arbitrary measurement function.  The 
decays of the Higgs are treated in the narrow width approximation.
At present, only the Higgs decay into two photons is included; 
other decays can be added. 
This Section provides instructions for using the code. 

\subsection{Download and Compile}
The code can be downloaded from~\cite{codeloc}. To compile the code, first uncompress the file 
{\tt fehip.tar.gz} and run the ``make'' script:
\begin{verbatim}
        \home\user> tar -zxvf fehip.tar.gz
        \home\user> cd FEHIP
        \home\user\FEHIP> make 
\end{verbatim} 
We have sussesfully compiled the code on Linux systems using the 
{\tt GNU g77} compiler. To compile the code on 
other platforms the user should modify the file {\tt makefile}.  
To run the code, execute the program {\tt fehip}:
\begin{verbatim}
        \home\user\FEHIP> fehip
\end{verbatim} 

The program performs multidimensional numerical integrations using the 
adaptive Monte Carlo algorithm Vegas~\cite{vegas};
we use its  recent implementation  in the  {\sf CUBA} library~\cite{cuba}. 
The current 
version of this library, {\tt Cuba-1.0}, is included in the distribution 
of \hname, and compiles automatically with the above {\tt make} script. 
For future updates of the {\tt Cuba} 
library, we refer the user to~\cite{cuba}; the 
directory {\tt FEHiP$\backslash$Cuba-1.0$\backslash$} and the 
file {\tt makefile} should be updated appropriately. 
 
\subsection{Basic parameters and Usage}

The basic  parameters that are used by \hname ~include 
the mass of the Higgs boson, the type and energy of the hadron collider, 
the factorization and renormalization scales, and the order in perturbation 
theory through which the result is to be computed.
The code also requires
the mass of the top quark and the value of the Fermi constant to compute 
the normalization of the Higgs boson production cross section
(cf. Eq.(\ref{eq85})). 
This input is provided by the user in the file  
{\tt input.txt}. 

A prototype input file has the following format:
\begin{verbatim}
'Mass of the Higgs boson (GeV) = ' 120d0
=============================================
'Collider (pp=0, ppbar=1)      = ' 0
'CMS collision energy (GeV)    = ' 14000d0
=============================================
'Factorization scale  (GeV)    = ' 120d0
'Renormalization scale  (GeV)  = ' 120d0
=============================================
'Top-quark mass (GeV)          = ' 175d0
'Fermi constant  =             = ' 0.0000116639d0
=============================================
'Final-state (0=no-decay, 1=diphoton) = ' 1
'Branching ratio                      = ' 1d0
=============================================
'Perturb. Order (0=LO, 1=NLO, 2=NNLO) = ' 1
=============================================
'Output File                   ='  'result.dat'
\end{verbatim}
The user should provide values for the following variables:
\begin{itemize}
\item {\bf The mass of the Higgs boson}. This is a {\it double precision} 
variable that  sets the value of the Higgs boson mass $m_h$, in 
GeV. 
\item {\bf Collider}. An {\it integer} variable that defines the 
type of collider. For proton-proton collisions, the value must  be set 
to {\bf 0}; for proton-antiproton collisions, the value must be set to 
{\bf 1}. 
\item {\bf Energy}. A {\it double precision} variable for the 
energy of the collider, in GeV.
\item {\bf Factorization scale}. A {\it double precision} variable for 
the value of the factorization scale $\mu_F$, in GeV.
\item {\bf Renormalization scale}. A {\it double precision} variable for 
the value of the renormalization scale $\mu_R$, in GeV.
\item {\bf Top-quark mass}. A {\it double precision} variable for the 
value of the top-quark mass $m_t$, in GeV.
\item {\bf Fermi-constant}. A {\it double precision} variable for the value 
of the  Fermi constant $G_F$, in $\mbox{\rm GeV}^{-2}$.
\item {\bf Final-state}. An {\it integer} variable that  determines 
the decay mode of the Higgs boson. The   value {\bf 0} corresponds to the 
``no-decay'' option. In this case 
it is assumed that the Higgs boson four momentum 
is fully reconstructed and is therefore an observable. 
With the ``no-decay'' option, it is possible 
to impose cuts on the transverse momentum and the rapidity of the 
Higgs boson, and to study jet-clustering and jet-veto cuts for the additional 
radiation (of up to $2$ partons) in the final-state.  However, no information 
about the kinematics of the Higgs decay product is provided.

The value  {\bf 1} corresponds to the 
``diphoton decay mode''. 
In this case, the Higgs decays into two photons in the narrow width 
approximation, and the kinematics of the two photons in the final 
state is fully reconstructed. Hence, arbitrary cuts on the photon 
momenta can be imposed.

We note that the results for the ``no-decay'' mode can always be obtained 
from the ``decay'' mode. Indeed, if 
no restrictions on the momenta of the individual photons are imposed, 
the two options should provide identical results. However, if one is not 
interested in the kinematics of the Higgs decay product, it is 
beneficial to declare the ``no-decay'' option explicitly, since the 
dimensionality of the numerical integration performed is then lowered.

Finally, we note that in the future, 
additional decays of the Higgs boson will be added to the program. 
 
\item {\bf Branching ratio}. A {\it double precision} variable 
for the value 
of the branching fraction of the Higgs boson decay into 
the selected final state. The program computes 
the Higgs production cross section in the narrow-width approximation
and  multiplies the output by the `Branching ratio'.
The value of the branching fraction for a particular decay of the Higgs  
with a certain mass should be supplied by the user as an input. Theoretical 
predictions for the branching fractions of the Higgs can be 
obtained with the program {\sf HDECAY}~\cite{hdecay}.

\item {\bf Perturbative Order}. An {\it integer} variable which sets the order 
through which the  perturbative expansion of the Higgs boson 
production cross section is computed. 
The values {\bf 0}, {\bf 1}, and {\bf 2}  
correspond to the LO,  NLO, and NNLO cross sections, 
respectively.  

\item {\bf Output File}. A {\it character string} variable for the file name 
where the results of the calculation are written.
The output file has the following format:
\begin{verbatim}
 proton-proton collider
 CMS collision Energy =   14000.
 ==================================
 Higgs boson mass =   120.
 Factorization scale =   120.
 Renormalization scale =   120.
 ==================================
 Top-quark mass =   175.
 Fermi constant =   1.16639E-05
 ==================================
 H --> gamma gamma
 Branching Ratio =   1.
 ==================================
 NLO Cross Section
 Strong coupling =   0.114230586
 ==================================
 
 ===========   RESULT    =================
 sigma =  38.4079658  +/-    0.0362102225
 chi square probability =  2.46758092E-05
 =========================================
\end{verbatim}
It contains a description of the input parameters and the result ({\tt sigma}) 
for the cross section in {\it picobarns} with the corresponding 
statistical error from the Vegas integration. It also provides 
the $\chi^2-$probability ({\tt prob} variable from the Vegas routine 
in {\sf CUBA}), which is  an indicator of the reliability of the 
Monte-Carlo integration.
\end{itemize}

 A number of other necessary program settings and options are not included 
in the user controlled {\tt input.txt} file. Some important settings 
can be found in the main program file {\tt higgs.F}; we have set 
them to values that are reasonable for practical purposes, according to 
our own experience with the code. In this file, we define the parameters 
for {\tt Vegas}, compute the normalization of the cross section, 
and invoke the integration routine.  For example, we require a precision 
(the {\tt epsrel} variable) of $1\%$ at NNLO and of $0.1\%$ at lower orders 
with an absolute error above $[10^{-3} \times Br]\;pb$.  We note that \hname~ 
also writes the iteration-by-iteration results produced by {\tt Vegas} to screen, 
so the user can track the output; this can easily be redirected to a separate file.  
This can be suppressed by modifying the appropriate flag in 
{\tt higgs.F}; we refer the reader to the {\sf CUBA} documentation for a 
discussion of the various output options.

\hname ~uses the MRST 2001 LO, NLO, and NNLO parton distribution 
functions~\cite{mrst}. These are provided with the distribution 
of \hname . We note that it is  extremely inefficient
to directly  call the pdf routines.  Since the factorization scale and the 
range of the Bjorken variable $x$ are fixed during each run of the program, 
we only need a limited amount of information about the pdfs. To optimize the code, we adopt the following 
strategy. When the code is initialized, \hname ~first calls the pdf 
routines for a given value of the factorization scale and computes 
the pdfs for $1000$ different choices of $x$. The distribution 
of these points is not uniform; we increase their density for $x$ values where the 
pdfs change rapidly. We then use interpolation 
with quadratic polynomials to connect the adjacent points. 
Consequently, each pdf is described by an array of coefficients of these 
quadratic polynomials which is kept in   
memory; this speeds up the execution of \hname ~enormously.  We have tested that the approximate 
pdf values are accurate to better than $0.05\%$ for all $x$ and $\mu_F$ values relevant for 
Higgs production.

To use a different set of pdfs, the user should 
replace the {\tt mrstpdfs.F} file with the appropriate set, and modify 
the calls to the {\tt mrstlo}, {\tt mrst2001}, 
and {\tt mrstnnlo} subroutines in the file {\tt fitpdf.F}.  In addition, 
the user should change the value of $\alpha_s\left(M_Z \right)$, 
which is used as an initial condition for the LO ,NLO, 
and NNLO evolution of the strong coupling, to the value which is 
consistent with the fitting of the new pdf set. 
The variables {\tt asZlo}, {\tt asZnlo}, and {\tt asZnnlo} 
in the file {\tt higgs.F} correspond to the fitted values for the
MRST 2001 LO, NLO, and NNLO  pdfs.

\subsection{Experimental cuts}

The most complicated user input is the definition of the experimental 
observable; this requires imposing cuts on the phase-space of the 
final state. 
We describe here a {\it double precision} function, named 
{\tt constraint}, in the {\tt constraint.F} file, which can be used for 
this purpose. 

The {\it constraint} function corresponds to the observable function 
$F_J$ for the final state.  The NNLO parton multiplicity is assumed:
\begin{equation}
\mbox{parton}_1(p_1)+ \mbox{parton}_2(p_2) \to 
H(p_h) + \mbox{parton}_3(p_3)+ \mbox{parton}_4(p_4).
\end{equation}
All final-state configurations with lower multiplicities have been mapped to 
this ``maximal'' final state by introducing additional partons with zero 
momentum when needed; this allows a uniform introduction of cuts for both real 
and virtual corrections. 

The routine uses the following (local and global) arguments : 
\begin{itemize}
\item the Vegas-generated variables: {\tt var};
\item the dimensionality {\tt ndim} of the Vegas integration;
\item the squared mass of the Higgs boson, {\tt m2} $= m_h^2$, and the 
CMS square energy of the colliding hadrons {\tt scm} $= s$;
\item the Bjorken variables {\tt x1}, {\tt x2} and the ratio 
{\tt z} $= \frac{m_h^2}{s x_1 x_2}$; 
\item the variables {\tt s13,s23,s14,s24} with
$ s_{if} = -\frac{p_i \cdot p_f}{ p_1 \cdot p_2}, \; i=1,2, \; f=3,4$,
and the variable {\tt s34} $= \frac{p_3 \cdot p_4}{ p_1 \cdot p_2}$;
\item the variables {\tt s1v,s2v} 
with  $s_{iv} = \frac{(p_i-p_h)^2}{2 p_1\cdot p_2}, \; i=1,2$; 
\item the variables {\tt s3v,s4v}  with $s_{vf} = \frac{(p_f+p_h)^2}{2 p_1\cdot p_2}, \; f=3,4$. 
\end{itemize}
From these variables we can reconstruct the components of the momenta of all
particles in the final state, and impose cuts which define the observable. 
We can also generate (using the {\tt var} array) the required additional 
phase-space variables for the subsequent decay of the Higgs boson.

After the appropriate variable declarations at the beginning of the routine, 
we proceed by generating phase-space variables for the two photons 
coming from the decay 
\begin{equation}
H(p_h) \to \gamma_a(q_a)+ \gamma_b(q_b),
\end{equation}
using a convenient parameterization in terms of invariant 
masses formed from the photon momenta $q_{a,b}$ 
and the initial state partonic momenta $p_{1,2}$.  
The routine, for each Vegas event, computes the following 
quantities. 
\begin{itemize}
\item The {\bf transverse energy} of: (1) the Higgs boson, {\tt pT};
(2) parton 3, {\tt pt3}; (3) parton 4, {\tt pt4}; (4) $\gamma_a$, {\tt ptga}; 
(5) $\gamma_b$, {\tt ptgb}. 
\item The {\bf energy} of: (1) the Higgs boson, {\tt En};
(2) parton 3, {\tt En3}; (3) parton 4, {\tt En4}; (4) $\gamma_a$, {\tt Ena}; 
(5) $\gamma_b$, {\tt Enb}.  
\item The {\bf momentum along the beam axis} of: (1) the Higgs boson, {\tt pZ};
(2) parton 3, {\tt pz3}; (3) parton 4, {\tt pz4}; (4) $\gamma_a$, {\tt pza}; 
(5) $\gamma_b$, {\tt pzb}.  
\end{itemize}

The above variables can be defined for all events, including the ones where
unresolved (soft/collinear) partons and photons 
are generated in the final state. However, the pseudorapidity and 
relative angles of massless particles can be sensibly defined only if 
they are resolved. To discriminate, we introduce a cutoff for the transverse 
energy of photons and partons in the final state, 
{\tt ptbuf $=0.01 {\rm GeV}$}. We then use two flags, 
{\tt topflag} and {\tt photonflag}, to characterize the event according to 
how many partons and photons have a transverse energy above the cutoff.  
For every event we  define: 
\begin{itemize}
\item {} {\tt topflag = 0d0}: two unresolved partons 
in the final state, {\tt pt3,pt4 < ptbuf}.
\item {} {\tt topflag = 1d0}: one resolved and one unresolved 
parton in the final state, {\tt pt3 < ptbuf, pt4 > ptbuf} or
{\tt pt4 < ptbuf, pt3 > ptbuf}. For such an event we 
compute the additional variables: 
\begin{itemize}
\item the transverse energy of the resolved parton: {\tt ptr}.
\item the rapidity of the resolved parton: {\tt etar}.
\end{itemize}
\item {} {\tt topflag = 2d0}: two resolved partons 
{\tt pt3,pt4 > ptbuf},  and the Higgs boson with non-zero transverse 
energy, {\tt pT > 0}. For this event we also compute: 
\begin{itemize}
\item the pseudo-rapidity of parton 3, {\tt eta3}, and parton 4, 
{\tt eta4}. 
\item the azimuthal angle between parton 3 and parton 4: {\tt phi34}.
\end{itemize}
\item {} {\tt topflag = 3d0}: corresponds to the (rare) configuration 
with two resolved partons, {\tt pt3,pt4 > ptbuf},  and the Higgs boson with 
zero transverse energy, {\tt pT = 0}.  For this event we also compute: 
\begin{itemize}
\item the pseudo-rapidity of parton 3, {\tt eta3}, and parton 4, 
{\tt eta4}. 
\end{itemize}
\end{itemize}
We also define:
\begin{itemize}
\item {} {\tt photonflag = 0d0}: at least one unresolved 
photon in the final state, {\tt ptga < ptbuf} or {\tt ptgb < ptbuf}. 
\item {} {\tt photonflag = 1d0}: corresponds to two resolved photons 
{\tt ptga,ptgb > ptbuf},  and the Higgs boson with non-zero transverse 
energy {\tt pT > 0}. For this event we also compute: 
\begin{itemize}
\item the pseudo-rapidity of $\gamma_a$, {\tt etaa}, and $\gamma_b$, 
{\tt etab}. 
\end{itemize}
\item {} {\tt photonflag = 2d0}: two resolved photons {\tt pt3,pt4 > ptbuf},  
and the Higgs boson with 
zero transverse energy, {\tt pT = 0}.  For this event we also compute: 
\begin{itemize}
\item the pseudo-rapidity of $\gamma_a$, {\tt etaa}, and $\gamma_b$, 
{\tt etab}. 
\end{itemize}
\end{itemize}
In order to study the isolation of photons, the routine computes the 
relative azimuthal angles of resolved photons with resolved partons. 
This can be done for the following combinations of flag values.
\begin{itemize}
\item {\tt photonflag = 1d0,2d0} and {\tt topflag = 2d0}: The program computes 
 the azimuthal angles for the pairs 
\begin{itemize}
\item $\gamma_a, \mbox{parton}_3$: angle {\tt phi3a}.
\item $\gamma_b, \mbox{parton}_3$: angle {\tt phi3b}.
\item $\gamma_a, \mbox{parton}_4$: angle {\tt phi4a}.
\item $\gamma_b, \mbox{parton}_4$: angle {\tt phi4b}.
\end{itemize}
\item {\tt photonflag = 1d0} and {\tt topflag = 1d0}: The program computes 
 the azimuthal angles for the pairs 
\begin{itemize}
\item $\gamma_a, \mbox{resolved parton}$: angle {\tt phira}.
\item $\gamma_b, \mbox{resolved parton}$: angle {\tt phirb}.
\end{itemize}
\end{itemize}

The above variables completely describe the phase-space for Higgs production 
and decay into photons through NNLO, and can be used to implement 
standard experimental cuts.  We now describe the structure of {\tt constraint.F}, 
and indicate how the reader modify the file to include other cuts.

The section of {\tt constraint.F} marked as {\tt STEP 1} performs the 
generation of the photon phase-space.  {\tt STEP 2 } computes the variables 
we have just described.  We have placed  the cuts required for the numerical results of this paper 
in the section labeled {\tt STEP 3: IMPLEMENTATION OF CUTS}.  These are:
\begin{itemize}
\item the Higgs boson rapidity cut;
\item the Higgs boson $p_T$ cut;
\item the jet-veto; 
\item cuts on the pseudorapiditiy and $p_T$ of each photon;
\item an isolation veto on each photon;
\item a cut on the average $p_T$ of the photons;
\item a cut on the pseudorapidity difference of the photons.  
\end{itemize}
An event is accepted if the {\tt constraint} function returns the value 
{\bf 1d0}, and rejected when it returns {\bf 0d0}. Before any cut is applied, 
the value of the function is set to 1d0. Cuts are applied successively; each 
cut could potentially reject the event by setting the function value to 0d0. 
To control which cuts should be applied to the event, we use two flags: {\tt 
active, inactive}. The user can choose combinations of the above cuts 
by modifying the appropriate flags, as described in {\tt constraint.F}.

The cuts that are programmed in {\tt STEP 3} of the routine serve 
as guiding examples for the user. When programming a new cut, 
some care is 
required for constraints on variables which are only defined when a final 
state particle is resolved; the user should provide that the 
{\tt constraint} function returns an  appropriate value for the events 
where the probed variable is not defined.  
In the section labeled {\tt STEP 4} 
of the {\tt constraint.F} file, 
we have placed a ``general'' cut to guide the reader through this process.  By modifying 
this section, it is possible to impose any desired constraint on the final state.

\subsection{Other information}
To summarize, the user input for \hname ~is a set of parameters in the 
{\tt input.txt} file, and the {\it constraint} function in the 
{\tt constraint.F} file. To run the program, the user should compile using the 
{\it make} script and execute {\bf fehip}. The {\it make} script should be 
executed every time that the {\tt constraint.F} file is modified. 
 
We now describe some standard running times for the observables considered in this 
paper.  We measure run times using the number of Vegas evaluations required to 
reach a given precision; on a 3 GHz PC, $10^5$ evaluations takes 
approximately 1 hour.  The required number of evaluations for 
observables in the ``no-decay'' mode is typically fairly small.  To reach 
1\% precision on the fully inclusive NNLO cross section at the LHC, 
about $1.5-2 \times 10^5$ evaluations are needed.  To reach 1\% precision on 
a jet-vetoed cross section, about $8 \times 10^5$ evaluations are needed.  These 
numbers increase dramatically when the ``decay'' mode is used, and the 
standard cuts on the photons are imposed.  For 1\% precision on the 
$pp \rightarrow H+X\rightarrow \gamma\gamma+X$ cross section with 
$p_T$, $\eta$, and isolation cuts imposed, $7 \times 10^6$ evaluations are 
required.  Computing bins for photon distributions 
to 2\% or better precision usually takes over $10^7$ evaluations.

Clearly, some observables require long running times. If the program is interrupted, 
the user can restart it by executing {\tt fehip}. 
The CUBA library saves in a file the data of every Vegas iteration; this is 
used to restart the program from the last completed iteration. To 
force the program to restart from the first Vegas iteration, the user should 
type {\tt make clean} and then execute {\tt fehip}.

\section{Conclusions}

We have discussed  the calculation of the Higgs boson production
cross section  in hadronic collisions through NNLO in perturbative QCD. 
The kinematics of both the Higgs and the QCD radiation is kept exactly.  
This allows us to consider any decay of the Higgs, and 
impose arbitrary cuts on the final state.  
In this paper, we have focused on Higgs decays into two photons.
We stress that this calculation provides the first example of a fully 
realistic NNLO calculation of the Higgs signal in 
the di-photon channel, where the two photons in the 
final state satisfy all the selection criteria (cuts on photon 
pseudorapidities, transverse momenta, and geometric isolation 
from significant hadronic activity) used by the ATLAS and CMS collaborations. 

To perform this calculation, we used the method of handling 
double real radiation described by us in \cite{sector,Anastasiou:2004qd}.  
We have further developed the approach in this paper, and have 
described its application to Higgs hadroproduction in detail.

We have given a description of the {\tt FORTRAN} code \hname ~which 
we used to obtain all the numerical results presented in this paper. 
The code can be obtained from \cite{codeloc}. We hope that \hname ~will 
be used when state-of-the-art knowledge of the Higgs signal at the 
LHC is required. In particular, since arbitrary cuts on the two 
photons and accompaning hadronic radiation are allowed, the code 
can be used to optimize cuts to enhance the signal-to-background 
ratio.

It was possible to obtain the results reported in this paper by further 
developing the approach suggested in \cite{sector,Anastasiou:2004qd}.
Since this computation is far more complex than those considered in \cite{sector,Anastasiou:2004qd}, 
we believe that this is an important milestone for our approach.  
However, further development of the method is desirable.  We give 
below a discussion from this perspective.  We indicate areas where 
more work is required, and describe possible solutions to the 
problems of the method.  We begin by listing attractive features of this approach.

\begin{itemize}

\item Given a
convenient parameterization of the phase-space for the real 
emission contribution to a given process, the singularities are 
extracted in an automated fashion; the human intervention 
required to achieve the final result is minimal.  No classification of the 
various singular regions is required.  Similarly, the 
cancellation of singularities is performed completely numerically, 
without the need for any analytic integrations.  In principle, this method 
therefore provides an algorithm for the extraction and cancellation of 
singularities through the ${\rm N^{n}LO}$ order in perturbation theory.  

\item Not only ``real'' singularities, but also some integrable 
ones are written in a factorized form. Consider the 
integral
\be
I_1 = \int \limits_{0}^{1} \frac{{\rm d}x {\rm d}y~x^{\ep}}{x+y}.~~~
\ee
Although the singularity at the point $x \sim y \sim 0$ 
in $I_1$ is integrable, we can apply the algorithm described in Section VII to it anyway.  
$I_1$ then becomes
\be
I_1 = I_1^{(a)} + I_1^{(b)},~~~~
I_1^{a} = \int \limits_{0}^{1} \frac{{\rm d}x {\rm d}y~x^{\ep}}{1+y},~~~
I_1^{b} = \int \limits_{0}^{1} \frac{{\rm d}x {\rm d}y~(xy)^{\ep}}{1+x}.
\ee
It is clear that the integrals $I_1^{(a,b)}$ are now written in forms very 
convenient for numerical evaluation. In addition to extracting real singularities, 
our technqiue also smooths integrable singularities, thereby improving the 
numerical behavior of the integrand.

\item The complete kinematic information of the process is preserved.  
In principle, this enables us to use this result 
to construct a partonic level Monte-Carlo event generator. 

\end{itemize} 

In spite of these attractive features, there are also some 
problems with this approach. 

\begin{itemize} 

\item 
The most important drawback of the method is that it
 produces large expressions, which 
require  long run-times for numerical 
evaluation  and lead to difficulties with optimizing 
the code.  Unfortunately, this is a natural feature of the 
approach; as we pointed out earlier, each time an entangled 
or line singularity is extracted, the expression size increases.  
However, it is useful to think of ways to ameliorate this behavior.  
It is instructive to 
recall how a somewhat similar problem was solved for NLO 
calculations.  It was found that squaring  matrix 
elements for unpolarized initial and final states is not 
a very practical option, since huge expressions are produced.  
Very compact expressions for scattering 
amplitudes can be obtained by working in the helicity basis.  
A similar approach can be attempted here.  However, a given 
helicity amplitude contains terms that belong to different 
topologies, in the language of this paper.  Since the separation into 
topologies is a crucial element of our approach, 
it is not clear how to to use helicity amplitudes efficiently.

\item 
By default, this approach might give more information than desired.  
All possible contributions to the cross section are automatically 
obtained.  For example, in the case of Higgs hadroproduction, 
we calculate $H + 2~{\rm jets}$ at LO, $H+1~{\rm jet}$ at
NLO and $H + 0~{\rm jets}$ at NNLO.  Imagine now that we want 
just the cross section of $H + 1~{\rm jet}$ at NLO.  Obviously, 
the number of entangled and line singularities we must deal 
with should be smaller.  However, a blind application of the 
algorithm in Section VII will automatically handle {\it all} 
singularities, including those needed for $H + 0~{\rm jets}$ at NNLO.  
It should be possible to implement additional criteria in the 
algorithm to prevent this from happening; for example, a 
check on the Higgs $p_T$ could be included, or a counting of the 
number of jets could be imposed.  In the calculation of 
$e^+e^- \rightarrow 2$ jets at NNLO described in \cite{sector,Anastasiou:2004qd}, 
the $e^+e^- \rightarrow 3$ jets cross section at NLO can be 
obtained efficiently by including a call to the jet algorithm in the 
routine that handles singularities.  However, a better 
understanding of these restrictions is needed before the approach can be 
used efficiently for processes with more complicated final states.

\item Making a partonic level event generator out of our result 
is tedious, although possible.
Because each sector corresponds to a different 
mapping of the invariant masses into $\lambda$-variables, 
each sector must be generated as a separate channel in a 
multi-channel Monte-Carlo. Our result 
for the Higgs hadroproduction cross section contains
approximately a hundred sectors.  In principle, this 
is not a problem, but it can definitely  become an important 
issue in practice. 
The only solution to this issue that we can see is a better choice of 
phase-space parameterization, which leads to a smaller number 
of sectors.

\end{itemize}

{\bf Acknowledgments.} We thank T. Hahn for helpful communications regarding
the numerical integration package CUBA. 
We are grateful to Michael Peskin and Zoltan Kunszt for their help in finding
the necessary computing resources, and the SLAC IT group for their technical 
support. We would like to thank John Campbell and Massimiliano Grazzini for 
comparisons. We are grateful to Guenther Dissertori and Zoltan  Kunszt for 
useful suggestions and discussions. We would like to thank Michael Dittmar
and Lance Dixon for pointing out mistakes in Figs.~\ref{fig:scale_cuts} and~\ref{fig:etadiff} in an earlier version of the paper. 
This work was started when the authors 
were  visiting the Kavli Institute for Theoretical Physics, UC California,
Santa Barbara. This research was supported by the
US Department of Energy under contract 
DE-FG03-94ER-40833 and the Outstanding
Junior Investigator Award DE-FG03-94ER-40833, and
by the National Science Foundation under contracts
P420D3620414350, P420D3620434350.


\end{document}